\documentclass[aps,prd,floatfix,10pt,nofootinbib,notitlepage]{revtex4-1} 

\usepackage{appendix}
\usepackage{hyperref}
\usepackage[margin=10pt,labelfont=bf, labelsep=endash, justification=centerfirst]{caption} 
\hypersetup{
pdfpagemode=UseOutlines,      
pdfstartview=Fit,             
pdffitwindow=true,            
pdfpagelayout=TwoColumnsRight,
pdftoolbar=true,              
pdfmenubar=true,              
bookmarksopen=false,          
bookmarksnumbered=true,       
colorlinks=false,              
pdfauthor={Florent Michel},     
pdftitle={},    
pdfcreator=PDFLaTeX,          %
pdfproducer=PDFLaTeX,         %
linkcolor=blue,               
urlcolor=blue,                
anchorcolor=black,            
citecolor=black,              
frenchlinks=false,             
pdfborder={0 0 0}             
}

\newcommand{\nn}{\nonumber\\}
\newcommand{\be}{\begin{eqnarray}} 
\newcommand{\ee}{\end{eqnarray}}
\newcommand{\om}{\ensuremath{\omega}}

\newcommand{\pd}{\ensuremath{\partial}}

\newcommand{\lp}{\ensuremath{\left(}}
\newcommand{\rp}{\ensuremath{\right)}}
\newcommand{\fp}{\ensuremath{f_{1,p}}}
\newcommand{\fb}{\ensuremath{f_{2,b}}}

\newcommand{\symbnode}{\ensuremath{n}}

\newcommand{\eq}[1]{Eq.~\eqref{#1}}

\newcommand{\beq} {\begin{equation}}
\newcommand{\eeq} {\end{equation}}
\newcommand{\bsub}{\begin{subequations}}
\newcommand{\esub}{\end{subequations}}
\newcommand{\bi} {\begin{itemize}}
\newcommand{\ei} {\end{itemize}}
\newcommand{\ben} {\begin{enumerate}}
\newcommand{\een} {\end{enumerate}}
\newcommand{\bmat} {\begin{pmatrix}}
\newcommand{\emat} {\end{pmatrix}} 
\newcommand{\bal} {\begin{aligned}}
\newcommand{\eal} {\end{aligned}}
\newcommand{\btab}{\begin{tabular}}
\newcommand{\etab}{\end{tabular}}

\usepackage{graphicx}		
\usepackage{subfig}
\usepackage{amsmath,amssymb,amsfonts,amsthm}
\usepackage{slashed}
\usepackage[utf8]{inputenc}   
\usepackage[T1]{fontenc} 
\usepackage{calligra}
\usepackage[french,english]{babel}
\usepackage{color}
\usepackage{amssymb}
\usepackage{mathrsfs}
\usepackage{pstricks}

   \usepackage{pstricks,pstricks-add}
   \usepackage{multido}

\begin{abstract}

To obtain the end-point evolution of the so-called black hole laser instability,we study the set of stationary solutions of the Gross-Pitaevskii equation for piecewise constant potentials which admit a homogeneous solution with a supersonic flow in the central region between two discontinuities. When the distance between them is larger than a critical value, we find that the homogeneous solution is unstable, and we identify the lowest energy state. We show that it can be viewed as determining the saturated value of the first (nodeless) complex frequency mode which drives the instability. We also classify the set of stationary solutions and establish their relation both with the set of complex frequency modes and with known soliton solutions. Finally, we adopt a procedure \textit{à la} Pitaevskii-Baym-Pethick to construct the effective functional which governs the transition from the homogeneous to nonhomogeneous solutions. 
\end{abstract}

\begin{document}
\selectlanguage{english}

\title{Saturation of black hole lasers in Bose-Einstein condensates}

\author{Florent Michel}
\email{michel@clipper.ens.fr}
\affiliation{Laboratoire de Physique Th\'eorique, CNRS UMR 8627, B\^atiment 210, Universit\'e Paris-Sud 11, 91405 Orsay Cedex, France}
\author{Renaud Parentani}
\email{renaud.parentani@th.u-psud.fr}
\affiliation{Laboratoire de Physique Th\'eorique, CNRS UMR 8627, B\^atiment 210, Universit\'e Paris-Sud 11, 91405 Orsay Cedex, France}

\date{\today}

\maketitle

\section{Introduction}
\label{intro}

The stability of inhomogeneous solutions of the Gross-Pitaevskii equation (GPE) 
is a rich and interesting topic, even when restricting one's attention, as we will do, to one-dimensional stationary flows which are asymptotically homogeneous. 
Among known stable solutions, one finds the so-called dark soliton~\cite{SolitonGPE,Pita-Stringbook,Goralectures} and another solution which is asymptotically divergent on one side \cite{belokolos1994algebro,GuidedBEC}.  
Both of these solutions will be used as building blocks of the solutions we will construct. 
It is clear that inhomogeneous flows that cross the speed of sound once can be dynamically stable but are necessarily energetically unstable since there always exist linear perturbations with negative energy.  
In addition, the mixing of these modes with the usual positive energy ones induces a super-radiance, which means that in quantum settings, there is a spontaneous production of pairs of phonons with opposite energy. Interestingly, this pair production is directly related to the Hawking prediction, according to which incipient black holes should spontaneously emit a thermal flux of radiation. This correspondence can be understood from the fact~\cite{unruhprl81,unruhprd95} that the curved space-time metric defined by a stationary flow that crosses the speed of sound once describes a black (or white) hole, the role of its event horizon being played by the supersonic transition.

When considering flows that cross the speed of sound twice, the phenomenology is even richer. 
In particular, it has been understood~\cite{BHL,BHLrevisited} that flows which 
are supersonic in a finite region and asymptotically 
homogeneous on both sides must be dynamically unstable because of a self-amplification of the super-radiance (the Hawking effect) occurring at each supersonic transition. It was then shown that the spectrum of linearized perturbations contains a discrete set of complex frequency modes which characterizes the dynamical instability~\cite{CoutantRP,FinazziParentani}. In fact, the supersonic region acts as an unstable resonant cavity, and the distance between the two "sonic horizons" governs the number of unstable modes. Below a certain value, there is no unstable mode and no pair production. In this case the flow is stable (dark). 
When increasing the distance, unstable modes appear one by one, each time with a higher number of nodes.
For large values, the number of unstable modes increases linearly with the distance.
In the present work, we complete the analysis in the particular case of piecewise-constant potentials such that the GPE admits a homogeneous solution with two sonic horizons. Similar configurations with a single horizon were considered in~\cite{gradino,2012PhRvA..85a3621L}. 
In addition, as done in~\cite{2regimesFinazzi} for a single horizon, 
we briefly show that the results obtained 
with the steplike approximation
apply to smooth profiles when the transition regions are sufficiently narrow.

Using the distance $2 L$ between the horizons as our control parameter, we first study 
the onset of the dynamical instability. We show that for a finite range of $L$, it is first described by an unstable mode with a purely imaginary frequency.\footnote{This fact was independently noticed by I. Carusotto, J.R.M. de Nova, and S. Finazzi (private communication).}
For larger distances, we recover the "normal" situation~\cite{CoutantRP,FinazziParentani} of complex frequency modes with properties directly linked to the Hawking effect. 
More precisely, each unstable degree of freedom first originates from a single quasi normal mode (QNM) 
when the latter frequency, which is purely imaginary,
 crosses the real axis. Then its frequency leaves the imaginary axis when a second QNM merges with it, so as to form a two-dimensional unstable system. 
These steps can be understood from the holomorphic properties of the determinant encoding the junction conditions across the two horizons which define the complex frequency modes. These properties severely restrict the conditions under which complex frequency modes can appear~\cite{Fullingbook}.

Second, following~\cite{GuidedBEC,Piazza,Baratoff,PhysRevB.53.6693}, 
we study the set of stationary nonlinear solutions of the GPE. 
We show that it is closely related to the discrete set of complex frequency modes which triggers the dynamical instability of the initial flow. Indeed, each unstable mode can be associated with a set of nine nonlinear solutions. In each set, the solution with the smallest energy may be conceived as the end point of the instability. Four of the nine solutions are smoothly connected to the homogeneous one, while the five additional ones 
contain either one or two solitons. When considering the full set of solutions for a given $L$, 
we show that the ground state of the system has no node, and contains no soliton. 
Finally, by a perturbative expansion of the GP energy functional similar to that used
by Pitaevskii~\cite{Pitaevskii} and Baym and Pethick~\cite{Baym} to study the spontaneous appearance of layered structures in flowing superfluids with a roton-maxon spectrum, we directly relate the set formed by the union of QNM and unstable modes to the above-mentioned four-dimensional subset of nonlinear solutions connected to the homogeneous one.

This paper is organized as follows. In Sec.~\ref{Slt}, we present the model, linearize the GPE, 
and find both the modes responsible for the dynamical instability, and the 
QNM from which they originate.
Exact stationary solutions of the GPE are studied in Sec.~\ref{stat_sol}, 
and their links with the linear solutions are given in Sec.~\ref{thermo}. Appendix~\ref{App:M-matrix} details the method we used to find complex frequency modes. Stationary solutions of the GPE with one single discontinuity are discussed in Appendix~\ref{App:single-h}.
Explicit formulas used to compute properties of solutions are given in Appendix~\ref{App:eqs_G_L}.

\section{Settings and linearized treatment}
\label{Slt}

\subsection{Settings}
\label{Settings}

We consider a one-dimensional flowing condensate with piecewise-constant two-body coupling $g$ and external potential $V$. We assume there are two discontinuities, located at $z=-L$ and $z=L$. We denote as $g_1,V_1$ the parameters in the left region, $I_1 : - \infty < z < -L$; $g_2,V_2$ the parameters in the central region, $I_2 : - L < z < L$; and $g_3,V_3$ the parameters in the right region $I_3 : L < z < \infty$. 
In each region $I_j$ ($j \in \lbrace 1,2,3 \rbrace$), the Gross-Pitaevskii equation reads
\be \label{GPE1}
i\frac{\partial \psi }{\partial t} = -\frac{1}{2}\frac{\partial ^2\psi }{\partial z^2}+V_j \psi +g_j \left|\psi ^2\right|\psi.
\ee
We work in units in which $\hbar$ and the atom mass are equal to unity. There is only one dimension in the problem, say the length, and time has the dimension of a length squared.
We consider stationary solutions, 
\be 
\psi (t,z)=e^{-i \mu t} f(z) e^{i \theta (z)},
\ee
where $f$ and $\theta$ are two real-valued functions, and $\mu \in \mathbb{R}$. Plugging this into 
\eq{GPE1} and using the definition of the (conserved) current, $J \equiv f^2 \pd_z \theta$,
we obtain
\be \label{eq:f}
f''= -2 \mu_j  f + 2 g_j f^3+\frac{J^2}{f^3},
\ee
where $\mu_j \equiv \mu - V_j$, and where a prime denotes differentiation with respect to $z$. 

We work with $g_j, \mu_j > 0$, and assume that the current is smaller than the critical value $J_{max}$, 
so that homogeneous solutions exist in each region [see Appendix~\ref{App:single-h}, \eq{eq:Jmax}]. 
We also assume that $g_j, \mu_j$ are such that there is a global homogeneous solution $f(z)=f_0$ 
with a subsonic flow in $I_1$, $I_3$, and a supersonic one in $I_2$. Hence this flow is a particular case of the black hole laser system studied in Refs.~\cite{BHL,BHLrevisited,CoutantRP,FinazziParentani}. 
Notice that the flow velocity $v$ is uniform in our homogeneous solution.   
To characterize the flow, it is convenient to work with 
\be \label{civ}
c_j^2 = g_j f_0^2,\, 
v = \frac{J}{f_0^2},
\ee
where $c_j$ is the sound speed in $I_j$, and $v = \pd_z \theta$ the constant condensate velocity.

\subsection{Complex-frequency modes}
\label{cfm}

The main properties of the set of unstable modes have been obtained by algebraic techniques in~\cite{CoutantRP}, and numerically in~\cite{FinazziParentani}. 
In particular these techniques were used to follow the evolution of the complex frequencies as a function of $L$. However, they were not able to describe the birth process of these modes when increasing $L$.
In the present settings, this can be analyzed in detail, revealing an interesting two-step process. In the body of the text we discuss the method and main results. 
The details of the calculation are presented in Appendix \ref{App:M-matrix}.

To obtain the equations for perturbations on the homogeneous solution $(f_0,\theta_0 (z)= z\, J/f_0^2)$,
we write $f(t,z)=f_0 + \delta f(t,z)$ and $\theta(t,z)=\theta_0(z)+\delta \theta(t,z)$, linearize \eq{GPE1}, and look for 
solutions of the form
\be \label{eq:fandtheta}
 \left\lbrace 
\begin{array}{ll}
 \delta  f_\om(t,z)=\Re \lp \delta F_\om \, e^{i (k_\om z - \omega  t)}  \rp &   \\
 \delta \theta_\om (t,z)=\Re \lp \delta  \Theta_\om \, e^{i (k_\om z - \omega  t)} \rp &.  
\end{array}
\right.
\ee
The linear equation gives
\be \label{eq:thetavsf}
\delta  \Theta_\om =2 i\lp \frac{v k_\om-\omega }{f_0 \, k_\om^2} \rp \delta  F_\om,
\ee 
and the dispersion relation
\be \label{eq:disprel1}
\Omega^2 \equiv (\omega -v k_\om)^2 = \frac{1}{4}k_\om^4 + c_j^2 k_\om^2 . 
\ee
At fixed frequency $\om$, 
\eq{eq:thetavsf} and \eq{eq:disprel1} characterize the four linearly independent solutions 
in each $I_j$. Solutions in different regions are related by matching conditions at $z = \pm L$ which follow from the continuity and differentiability of $f$ and $\theta$.

We are interested in computing the discrete set of complex-frequency modes, with eigenfrequencies
$\om_a \in \mathbb{C}-\mathbb{R}$, which trigger the laser effect.  
So, we should only consider asymptotically bounded 
modes (ABM), \textit{i.e.}, keep the waves $e^{i k_\om z}$ 
which decay exponentially as $z \rightarrow \pm \infty$~\cite{CoutantRP}.
In $I_1$ and $I_3$ there are two such solutions (for a given sign of $\Gamma \equiv \Im \om$). For $\Gamma > 0$, they correspond to the analytical continuations in $\om \in \mathbb{C}$  of the outgoing wave and the exponentially decreasing wave.
In the central region $I_2$, the four waves are kept. In the present case, eight boundary conditions must be satisfied (continuity and differentiability of $\delta f$ and $\delta \theta$ at $z=\pm L$). 
They impose eight linear relations between the coefficients of the waves,
which can be written as an 8-by-8 matrix $M(\om)$. This system has nontrivial solutions if 
$det M(\om)=0$, which selects the sought-for discrete set of frequencies. 

To study how these ABM appear as $L$ increases, 
we will consider a larger (still discrete) set which 
includes quasinormal modes (QNM) which are not asymptotically bound.
Using this larger set, we will see that every complex frequency ABM arises from two QNM in two steps. To understand the origin of these two steps, one should recall that in the general case, each dynamical instability is described by a two-dimensional system which corresponds to a complex unstable oscillator; see~\cite{Rossignoli} and Appendix C in~\cite{CoutantRP}. 
Such a system is composed of
two {\it complex} eigenmodes of frequency $\om_a = \Re \, \om_a  \pm  i\Gamma_a $, with $\Re \, \om_a, \Gamma_a  > 0$. Only the mode which grows in time is {\it outgoing}.  
By outgoing, we mean the following:  
the group velocity $v_g = (\partial_\om k_\om)^{-1}$ 
of the analytic continuation of the two roots $k_\om$ that are real  
for real $\om$ is pointing outwards. 
Besides this case, there also exists a degenerate case, not considered in~\cite{CoutantRP,FinazziParentani}, 
described by only two {\it real} modes with imaginary frequencies $\pm i \Gamma_a$~\cite{Rossignoli,Fullingbook}. 
In this case too, the ABM which grows in time is outgoing in the above sense. 
Interestingly, the two-step process we found is directly associated with this degenerate case.

When looking for QNM, we should also pay attention to the implementation of the 
outgoing boundary conditions because there are four roots in \eq{eq:disprel1}, not two as in the standard definition of QNM~\cite{reviewKokotas,QNM_BH_BB}.
We adopt the same definition as the one above determining the ABM:
We keep the analytical continuations in the complex lower half-plane of the outgoing 
wave and the exponentially decreasing wave for $\om \in \mathbb{R}$. 
With this definition, $det M = 0$ gives all outgoing modes, that is, 
the spatially ABM for $\Gamma > 0$, and the QNM for $\Gamma < 0$, 
both for the standard and the degenerate case with $\Re \om = 0$. 

\subsection{Results}
\label{res_ABM/QNM}

To study the two-step process for increasing values of $L$, we work with $c_3=c_1$, and then briefly discuss the changes when $c_3 \neq c_1$. 
We first find that every new ABM appears in the degenerate sector, at $\om = 0$, and for values of $L$ given by
\be \label{eq:Lm} 
L_m \equiv L_0 + \frac{\lambda_0}{2} m, 
\ee
where 
\be \label{eq:Lcrit}
L_0 &=&  \frac{1}{2 \sqrt{v^2 - c_2^2}} \arctan \lp \sqrt{\frac{c_1^2 - v^2}{v^2 - c_2^2}} \rp, 
\\
\label{eq:lambda}
\lambda_0 &=& \frac{\pi}{\sqrt{v^2 - c_2^2}},
\ee
and where $m \in \mathbb{N}$. This is the first step.
For each $m$, it is followed at $L = L_{m +1/2}$ by 
a merging process when the frequency of a QNM 
crosses the real line and equals that of the degenerate ABM. For $L > L_{m+1/2}$, the $m$th unstable sector is described by the nondegenerate case, i.e. by a complex ABM with a complex frequency. 

\begin{figure}
\includegraphics[scale=0.7]{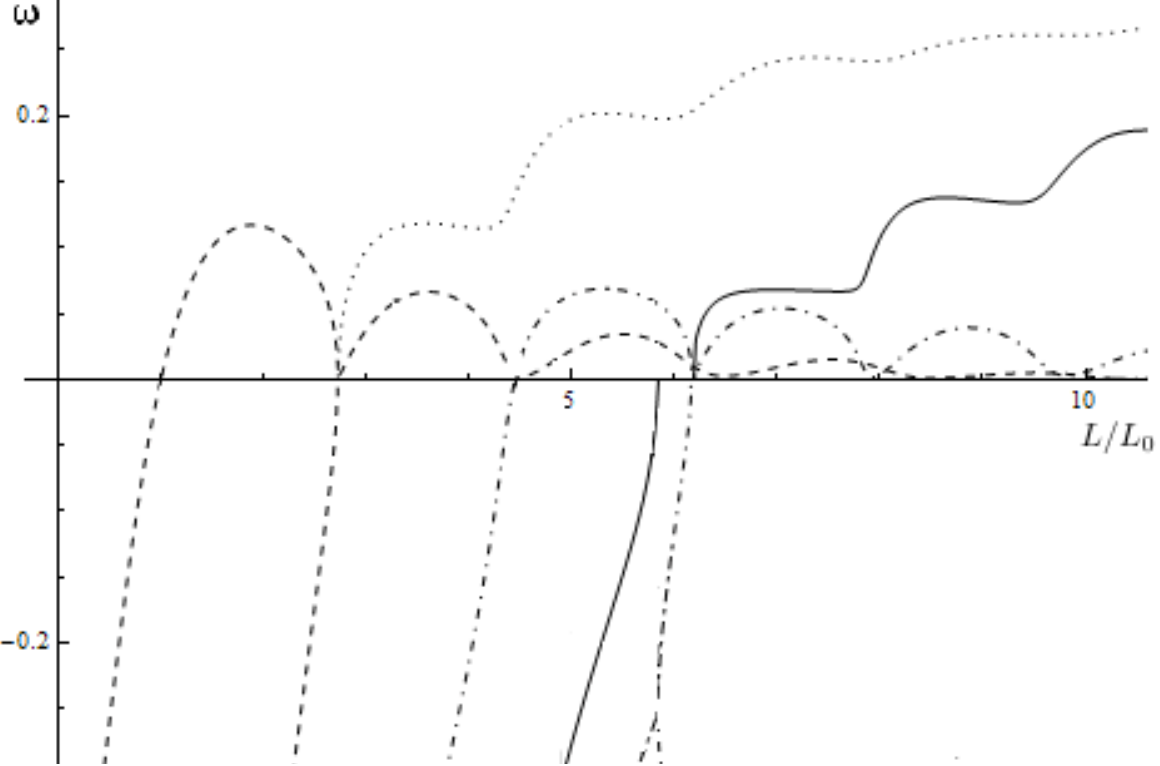}
\caption{Evolution of the two complex frequencies composing the first two sectors $\symbnode=0$ and $\symbnode=1$ as functions of $L/L_0$.  Dashed (dotted) line: Imaginary (real) part of the two frequencies associated with $\symbnode=0$. Solid (dash-dotted) line: 
Imaginary (real) part for $\symbnode=1$.
The solid line for negative values gives $\Re \om$
of the second QNM. At $L = L_0$ of \eq{eq:Lcrit}, the first ABM appears, as the QNM frequency crosses the real axis. 
The ABM frequency remains purely imaginary until $L = L_0 + \lambda_0/ 4 \sim 2.8 \, L_0$, where the second QNM of the first sector merges with it. 
For larger $L$, the frequency is complex (only the solution with $\Re \om > 0$ is represented). The story is similar for the second sector $\symbnode=1$. Notice that the second QNM frequency extends further than what is shown on the plot, as it starts at $\om \approx - i 6$. Note also that it has a complex frequency for small values of $L$. This complex QNM splits into two purely imaginary QNMs for some value $L_i$ close to $L_{3/2}$, as  can be seen in the bottom of the figure. 
The parameters are: $v=1.0$, $c_1=1.5$ and $c_2=0.5$.} \label{fig:QNM}
\end{figure}

Surprisingly, when restricting our attention to zero-frequency solutions, the critical values $L_{m}$ with $m$ 
{\it integer} or {\it half-integer} appear altogether. Indeed, linearizing \eq{eq:f} in $\delta f$ and assuming the solution is static and bounded at infinity, one gets
\be 
\delta f(z)=
\left\lbrace 
\begin{array}{ll}
 A_L \exp \left(2 \sqrt{c_1^2-v^2} \, z\right), & z<-L ,\\
 A \cos \left(2 \sqrt{v^2-c_2^2} \, z+\varphi \right), & -L<z<L, \\ 
 A_R \exp \left(-2 \sqrt{c_1^2-v^2} \, z\right), & z>L ,
\end{array}
\right. 
\ee
where $A_L$, $A$, $A_R$ and $\varphi$ are real constants. The matching conditions at $z=\pm L$ give
\be \label{eq:tan}
\tan \left(2 \sqrt{v^2-c_2^2} \, L+\varphi \right)=-\tan \left(-2 \sqrt{v^2-c_2^2} \, L+\varphi \right)=\sqrt{\frac{c_1^2-v^2}{v^2-c_2^2}}.
\ee
This implies that $\varphi =0\, \text{modulo} \, {\pi }/{2}$ 
 and $L$ obeys \eq{eq:Lm} with $m$ 
 integer or half-integer. 
At this level, it might seem that a new ABM is obtained for all these values of $L$. This is not quite correct, as is revealed by studying the solutions of \eq{eq:fandtheta} with $\Re \om \neq 0$, 
 see Fig.~\ref{fig:QNM} and Sec~\ref{thermo}.

When including the QNM, the picture gets clearer, as one can see that both steps occur
when a QNM frequency crosses the real axis.
Starting with $L=0$, we obtain the following sequence; see Fig.~\ref{fig:QNM}.
When $L = 0$, there is no ABM, but there is already one QNM.
This is the ancestor of the first ABM. Indeed, when $L$ increases, the frequency moves along the imaginary axis, and when it crosses the real axis, it becomes the first ABM. 
The onset of instability occurs at $L=L_0$ of \eq{eq:Lcrit}.
As shown in Sec.~\ref{stat_sol}, $L_0$ is also the value of $L$ at which the Gibbs energy of a nontrivial nonlinear solution becomes smaller than that of the homogeneous solution. 
As expected~\cite{Rossignoli}, the dynamical (linear) instability thus appears together with an energetic instability.   
When $v < c_2$, the homogeneous solution is everywhere subsonic and there is no dynamical instability. There are still QNM, but these never cross the real axis to become ABM. 

When further increasing $L$, the ABM frequency keeps moving along the imaginary axis.
$\Im \om$ reaches a maximum value $\Gamma_M$, and then starts to decrease, still along the imaginary axis. 
Besides this, a second QNM appears on the negative imaginary axis and moves up.
This new QNM merges with the ABM at $\Gamma=0$ for $L = L_{1/2}$.
For higher $L$, the ABM eigenfrequency leaves the imaginary axis. The evolution is then similar to what was found in \cite{FinazziParentani}. The imaginary part shows oscillations with a decreasing amplitude, while the real part goes to $\om_{max}$ given by \eq{eq:omMax}.
By a numerical analysis of $det M = 0$, we found couples of QNM for $\symbnode \equiv  \lfloor m \rfloor \in \left[0,8 \right]$.
As we will see in Sec.~\ref{stat_sol}, nonlinear solutions are classified by an integer
number $\symbnode$ which labels the harmonics in the central region.
Notice also that $\symbnode$ 
coincides with the Bohr-Sommerfeld number $n_{\rm BS}$ used in~\cite{FinazziParentani}; see Appendix~\ref{App:M-matrix}.
For each $n$, one QNM crosses the real axis with a vanishing real part at $L = L_n$, therefore becoming the new ABM. Then the latter merges with another QNM at $\Gamma=0$ for  $L = L_{n+1/2}$ before leaving the imaginary axis. The subsequent evolution is similar to the case $\symbnode=0$. 
We conjecture that this remains true for any $\symbnode \in \mathbb{N}$ since the stationary analysis giving 
\eq{eq:tan} applies to all $n$. The cases $\symbnode=0$ and $\symbnode=1$ are represented in Fig.~\ref{fig:QNM}. 

In all cases, we notice that QNM and ABM frequencies never leave the imaginary axis except when they merge with another one. This is due to the continuity and differentiability of $det M$ in $\om$, as well as its symmetry under $\om \rightarrow - \om^*$, $k \rightarrow - k^*$ (this holds if $\Gamma$ is larger than some critical value $\Gamma_c < 0$ , which is always the case for the imaginary modes we describe). Indeed, a mode leaving the imaginary axis must turn into two modes $\om$ and $-\om^*$. The change in the phase of $det M$ when turning around them in the complex $\om$ plane is then equal to $4 \pi$ times some integer. But turning around one single ABM (or QNM) frequency gives, in general, a change of phase of $\pm 2 \pi$ since $det M$ is linear close to it. So, by continuity of the phase of $M$, a frequency cannot leave the imaginary axis, except when two frequencies merge. 

When considering $c_1 \neq c_3$ we found the following. \eq{eq:Lm} remains true, with $\lambda_0$ still given by \eq{eq:lambda} and $L_0$ given by
\be \label{eq:lambda_c3neqc1} 
L^{c_1 \neq c_3}_0 =  
\frac{1}{4 \sqrt{v^2 - c_2^2}} \lp \arctan \lp \sqrt{\frac{c_1^2 - v^2}{v^2 - c_2^2}} \rp + \arctan \lp \sqrt{\frac{c_3^2 - v^2}{v^2 - c_2^2}} \rp \rp.
\ee 
One also finds that ABM have a finite imaginary part when they leave the imaginary axis, and the first QNM appears at a finite value of $L$. 

We end this section by noting that we observe in Fig.~\ref{fig:QNM} a strong parallelism between the curves followed by the QNM frequencies, especially the first two. This indicates that there might be an approximative discrete translation invariance. This is reinforced by the fact that the difference in $L$ between them is $\lambda_0/4$, which corresponds to a symmetry of $det M$ in the limit $\Gamma \rightarrow 0$. It is currently unclear to the authors whether this symmetry alone can explain the observed parallelism.

\section{Nonlinear stationary solutions}
\label{stat_sol}

In this section we describe exact solutions to the time-independent GPE in black hole laser configurations.
Our method is similar to that used in \cite{GuidedBEC} to describe a propagating Bose-Einstein condensate through a wave guide with an obstacle. Related ideas were also used in \cite{periodic_gradini}. We limit ourselves to solutions whose amplitudes go to $f_0$ at $z \rightarrow \pm \infty$ since only they have a finite energy. Our aim is to classify the set of solutions and to find the ground state of the system when the homogeneous configuration is unstable, \textit{i.e.}, for $L > L_0$. For simplicity, unless explicitly stated otherwise, we assume the microscopic parameters $g$ and $\mu$ are identical in $I_1$ and $I_3$: $ g_1 = g_3 , \, \mu_1 = \mu_3$. 

We use the Gibbs energy $E$ of \eq{eq:E}, which means that we work in the ensemble where the chemical potential $\mu$, the temperature (set to zero) and the current are fixed. In this ensemble, the system is characterized by the parameters $g_1$, $g_2$, $\mu_1$, $\mu_2$, $J$ and the interhorizon length $2 L$. They are not independent: The assumption that a globally uniform solution exists gives a relation between them since the two polynomials 
$2 g_j \, f^6- 2 \mu_j \, f^4 +J^2$ evaluated in regions $1$ and $2$ must have a common root $f_0$;
see \eq{poly1}. When setting $f_0$ to unity by a rescaling of the unit of length, we have
\be 
2 g_1 - 2 \mu_1  +J^2=2 g_2 - 2 \mu_2  +J^2=0. 
\ee 
The system depends only on four parameters, for instance $(c_1,c_2,v,L)$; see \eq{civ}. In the black hole laser case, we have $0 < c_2 < |v| < c_1$.
\begin{figure}
\begin{center}
\includegraphics[scale=0.9]{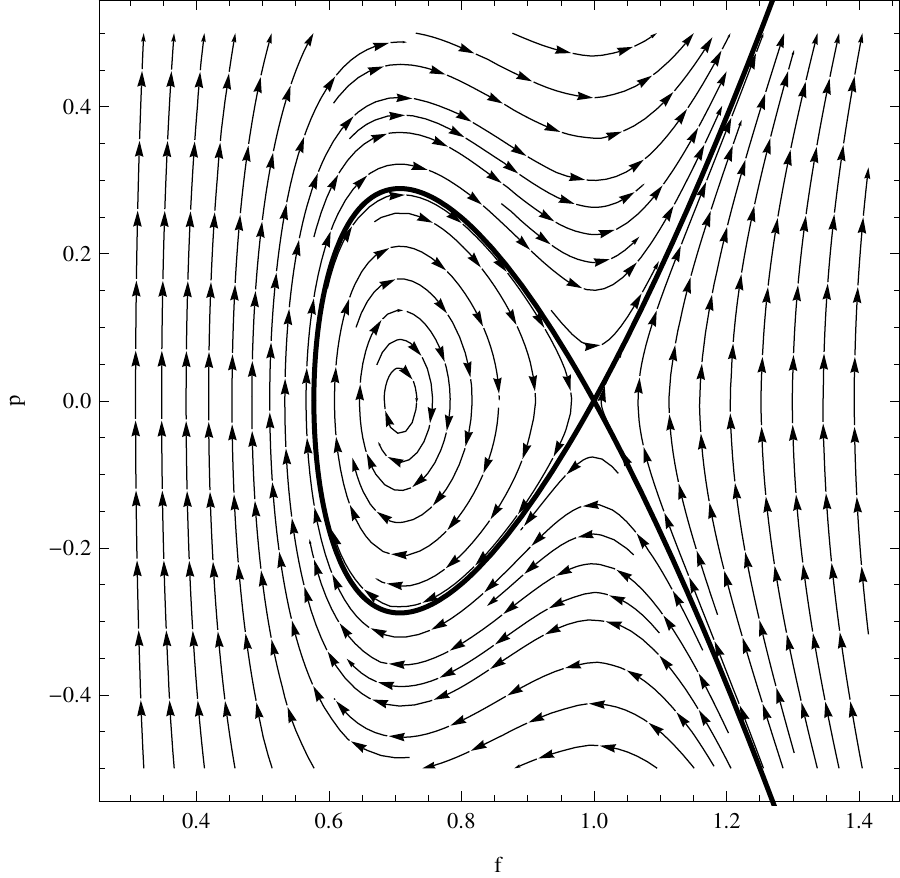}
\includegraphics[scale=0.9]{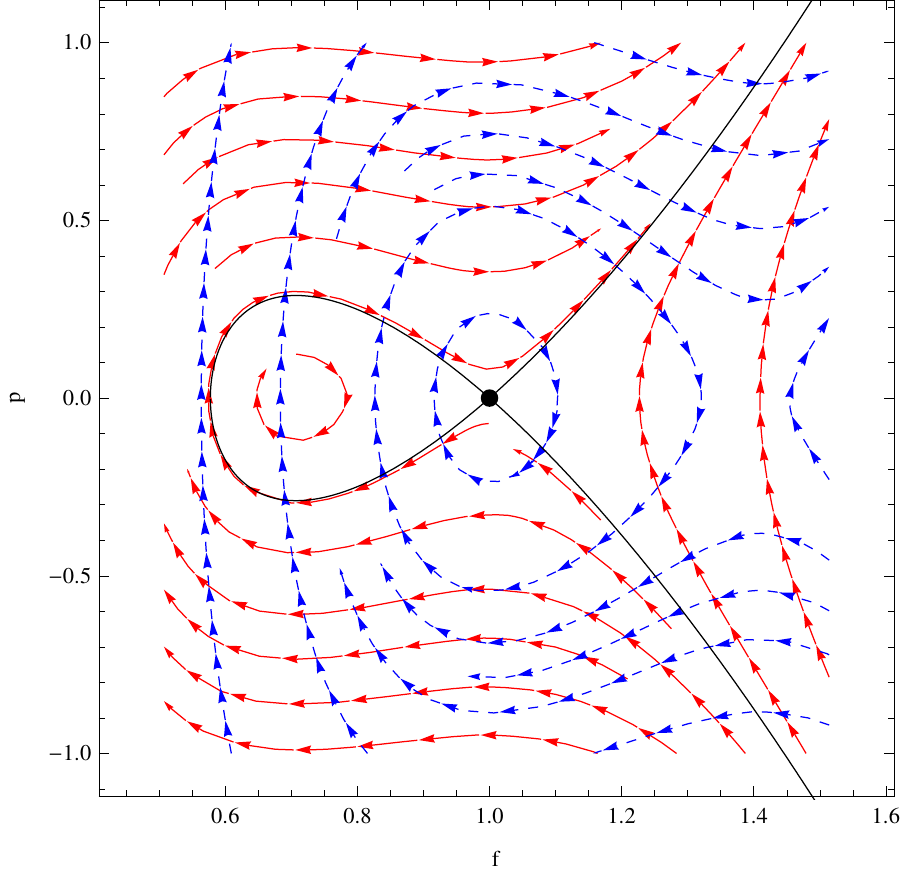}
\caption{Left panel: Phase portrait $p$ versus $f$ of \eq{keyE}.
 It contains three qualitatively different regions, separated by the thick lines. These lines correspond to solutions which go asymptotically to a finite value.
  The middle domain contains periodic bounded solutions. Solutions in the right and left domains are divergent at finite values of $z$.  
Right panel: Two superimposed phase portraits corresponding to regions $I_1, I_3$ (solid, red) and $I_2$ (dashed, blue). The black dot represents the globally homogeneous solution $f_0$, and the black lines are the solutions which reach $f_0$ at infinity. The parameters of both panels are $g_1=8$, $g_2=1$, $\mu_1=28/3$, $\mu_2=7/6$ and $J^2=8/3$.}\label{phase_portrait2}
\end{center}
\end{figure}

The main properties of the solutions can be seen on the phase portrait, which represents the trajectories of the solutions in the $(f,p=f')$ plane; see Fig.~\ref{phase_portrait2}, left panel. The key equation in $I_j$ is given by the integral of \eq{eq:f}, namely,
\be \label{keyE}
p^2 = f'^2=\frac{1}{f^2}\left(g_j f^6-2 \mu_j  f^4+C_j f^2-J^2\right),
\ee
where $C_j$ is the integration constant. 
Figure~\ref{phase_portrait2} right panel shows a superposition of the two phase portraits for the regions $I_1$ and $I_3$ (red,solid) and $I_2$ (blue, dashed).  Its qualitative properties, in particular the ordering of the three stationary points and the behavior of solutions around them, do not depend on the precise values of the parameters. They would change if we allowed $c_1<|v|$ or $c_2>|v|$. 

\begin{figure}
\begin{center}
\includegraphics[scale=1.0]{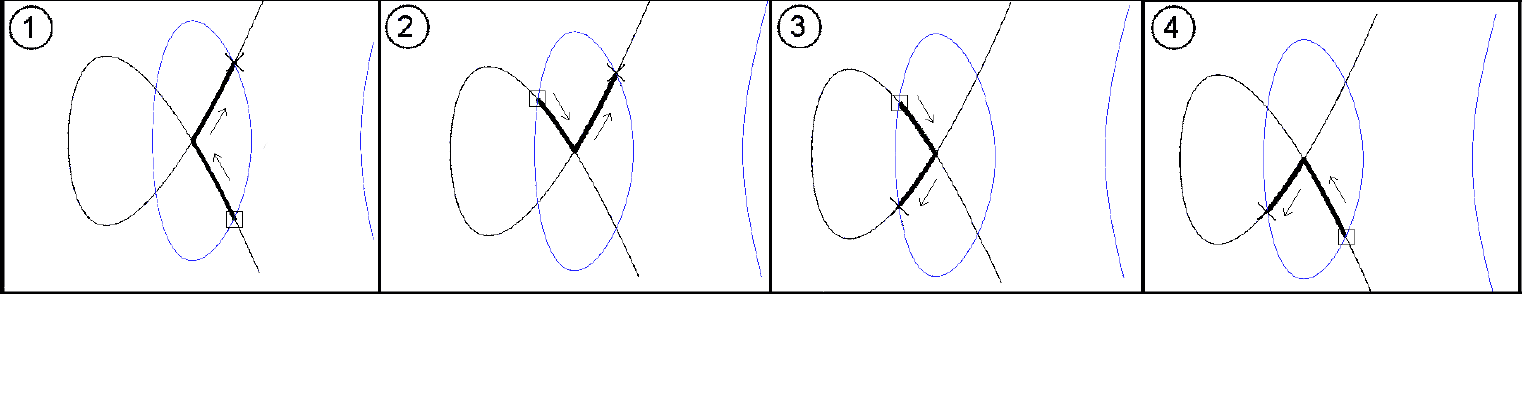}
\includegraphics[scale=1.0]{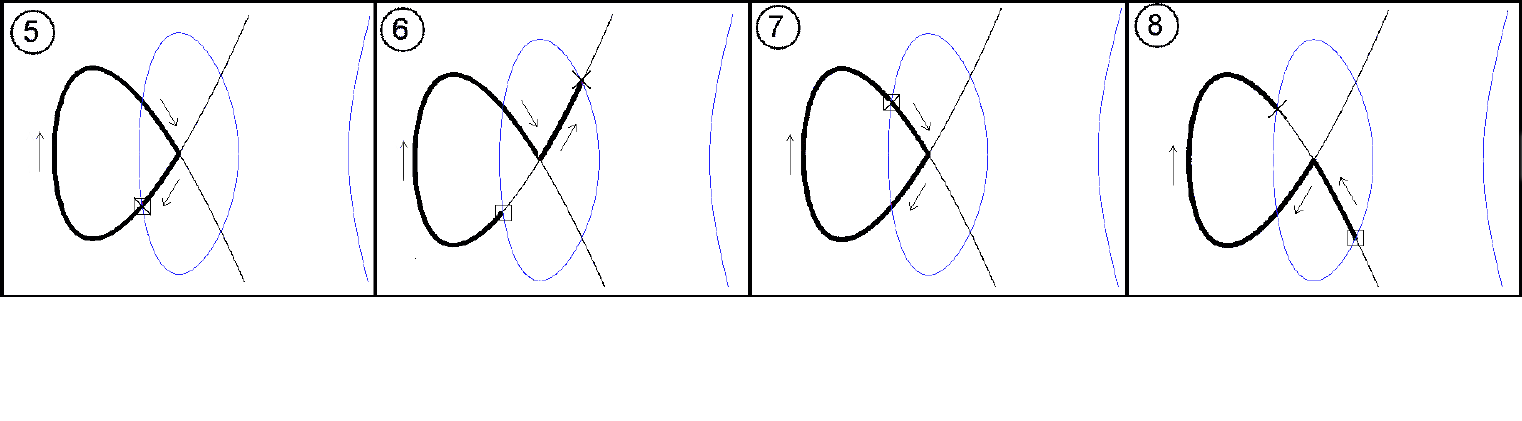}
\includegraphics[scale=0.3]{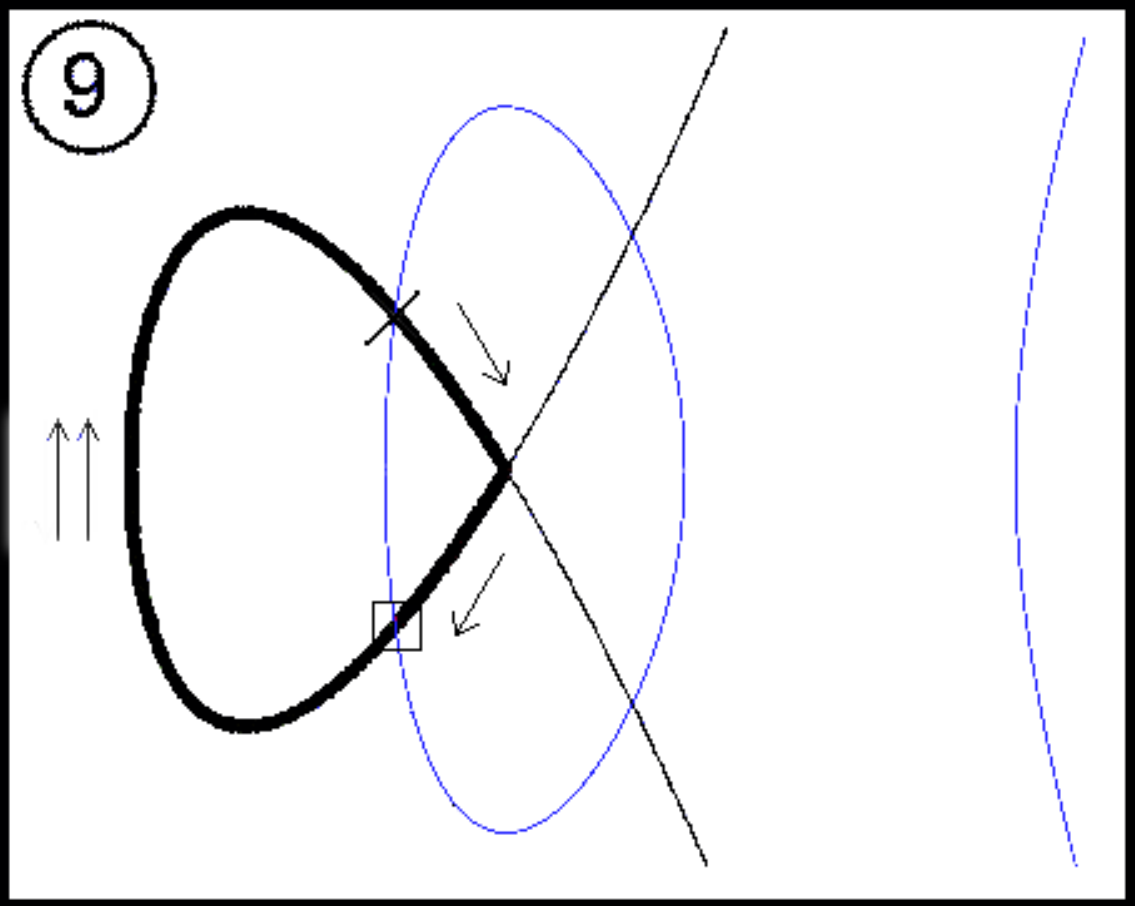}
\caption{The nine different types of trajectories in phase space: 
The first four solutions (top line) continuously connect to the homogeneous one, the next
four solutions have one soliton (middle), and the last solution (bottom) has two solitons.  
On each plot the black line of Fig.~\ref{phase_portrait2} is plotted along with the blue lines corresponding to a given value of the integration constant $C_2$. Thick black curves correspond to trajectories in phase space in $I_1$ and $I_3$, the direction being indicated by an arrow. The double arrow in the last plot indicates that a part of the curve is followed twice: once in $I_1$ and once in $I_3$. In $I_2$, the solution follows the closed blue line clockwise, starting from the first intersection with the thick one (materialized by a cross) at $z=-L$ and ending at the second intersection (box) at $z=L$. In between it can make an arbitrary number $n \in \mathbb{N}$ of turns.}\label{fig:9-traj}
\end{center}
\end{figure}

We are interested in solutions for which $f \rightarrow f_0$ as $z \rightarrow \pm \infty$. 
So, in Fig.~\ref{phase_portrait2} the solution must start on the black dot $f=f_0$, $f'=0$ at $z = - \infty$. When $z$ is increased the solution either remains at that point (for the globally homogeneous solution) or moves along the black line until $z=-L$. It then follows the flow of the blue dashed lines until $z=L$. Finally, for $z>L$ it follows a black line again up to the black dot, which it reaches asymptotically. As described in Appendix~\ref{App:single-h}, for a given value of the integration constant $C_2$, there are three possible trajectories in phase space for $z \in (-\infty,-L)$. The same is true for $z \in (L,\infty)$.
No restriction should be put \textit{a priori} on the solution in the central region $I_2$ since it is finite and will contribute to $E$ by a finite amount provided $f$ is regular in $I_2$. However, an inspection of the phase portrait in Fig.~\ref{phase_portrait2} reveals that, because of the matching conditions at $z = \pm L$, the solution in $I_2$ must lie in the central domain of the phase portrait in Fig.~\ref{phase_portrait2}. \footnote{In fact there exists one solution (type 3 in Fig \ref{fig:9-traj} for $n=0$) which can extend to values of $C_2$ giving solutions in the external domains. Whether it does so depends on the precise values of the parameters.}
As a result, the solution is characterized by the number of cycles in $I_2$, $n \in \mathbb{N}$, and the integration constant $C_2$. In total, for a given value of the discrete parameter $n \in \mathbb{N}$, there are nine different types of solutions. They are represented in Fig.~\ref{fig:9-traj}. Notice that for each of them, the value of the parameter $C_2$ is fixed by $L$. Also, the minimum value of $L$ at which solutions exist goes to infinity as $n \rightarrow \infty$. Hence, for a fixed $L$, there exists only a finite number of solutions. 

When $L$ is smaller than $L_0$ of \eq{eq:Lcrit}, only two solutions exist: the homogeneous one and 
another one of type 3 in Fig.~\ref{fig:9-traj} with $n=0$. As shown in \eq{eq:E-os}, the energy density change in $I_j$ (with respect to the homogeneous solution) is
\be 
\Delta\mathcal{E}_j = -\frac{1}{2}g_j \lp f^4-f_0^4 \rp - J^2 \lp \frac{1}{f^2} - \frac{1}{f_0^2} \rp .
\ee
For $L < L_0$, the nonuniform solution has a positive energy. 
Hence the homogeneous configuration is stable. 
When $L > L_0$, the inhomogeneous solution is replaced by that corresponding to plot 1 in Fig.~\ref{fig:9-traj}, which has a negative energy. 
Therefore the homogeneous solution becomes energetically unstable at $L=L_0$.
This confirms the results of our linear analysis presented in Sec.~\ref{Slt} where the first dynamical instability was found for $L > L_0$. So, as expected from \cite{Jackson,Rossignoli}, the dynamical instability appears together with a static instability when a solution becomes thermodynamically more favorable than the uniform one. Notice that the transition at $L=L_0$ is a second order one
since the amplitude of the oscillations in $I_2$ goes to zero as $L \rightarrow L_0$. Notice also that when $|v|<c_1,c_2$, the uniform solution is always stable, while if $|v|>c_1$ it is always unstable. 

Figure~\ref{fig:ene-9-eq} shows $\Delta E$ as a function of $L/L_0$. The formulas we used are presented in Appendix~\ref{App:eqs_G_L} [see Eqs. (\ref{eq:GtoE}-\ref{eqs-num-end})].
This figure first establishes that the type 1 solution with $n=0$ is indeed the lowest energy state. We also see that at large $L$, $\Delta E(L)$ becomes linear for all solutions with a negative slope $\frac{1}{2} g_2 (f_0^2 - \fb^2) + J \lp f_0^{-2} -\fb^{-2} \rp$, where $\fb$ is the subsonic uniform solution in $I_2$ given in \eq{eq:fb}.
Note that for $n \neq 0$ and $L$ slightly smaller than its critical values there are two solutions 
of type 3. This is because the length $L$ associated with this series of 
solutions is not monotonous in the integration constant $C_2$. It decreases close to its minimum value but then increases with $C_2$.

\begin{figure}
\begin{center}
\includegraphics[scale=0.5]{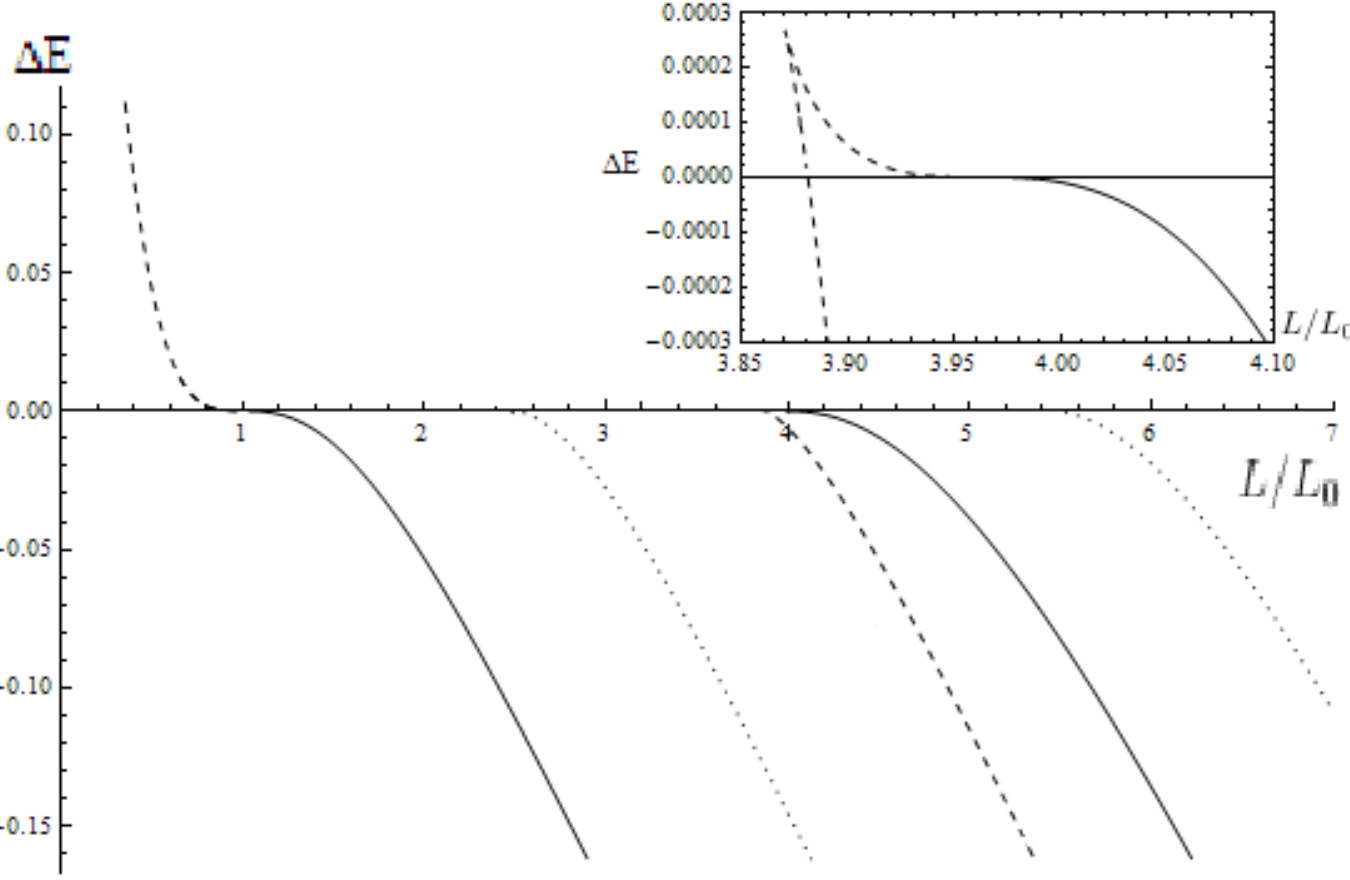} 
\includegraphics[scale=0.6]{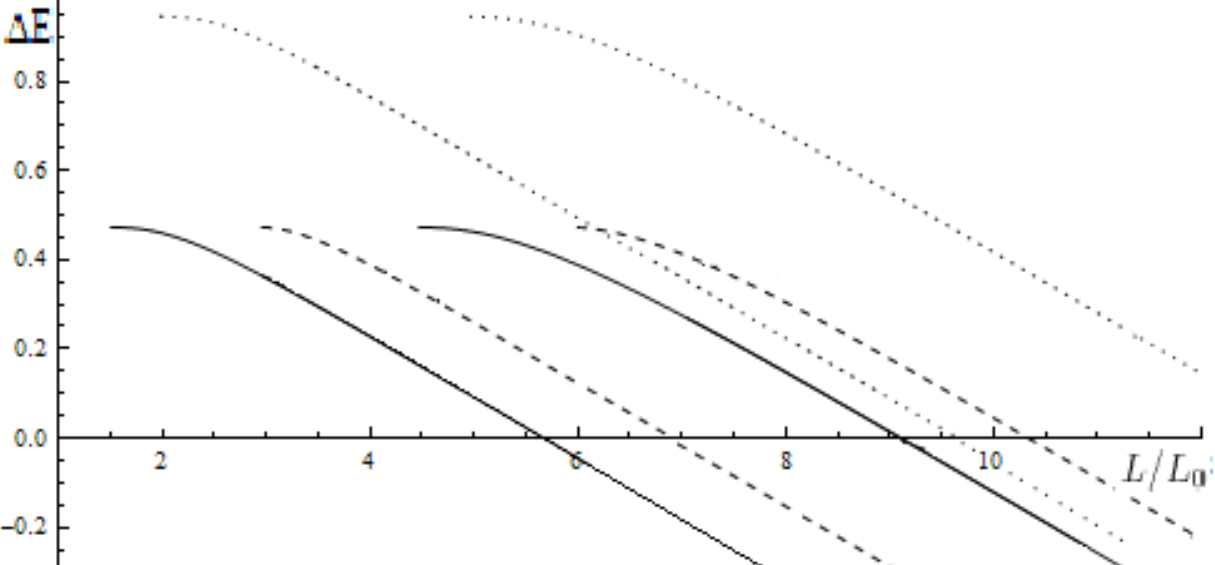} 
\end{center}
\caption{Left panel: Energy differences $\Delta E$ for the four different types of solutions with no soliton 
as functions of $L/L_0$. The number of cycles $n$ is equal to $0$ and $1$.
Solid lines: 
Type 1 in Fig.~\ref{fig:9-traj}; dotted lines: 
Types 2 and 4 (degenerate for $c_1 = c_3$),
and dashed lines: 
type 3. We set $c_3^2=c_1^2=8$, $c_2^2=1$, $v^2=8/3$ and $f_0=1$. 
The insert shows a zoomed-in picture of the beginning of the curve for types $1$ and $3$ when $n=1$.
As explained in the text, for $n= 0$ type 3 exists from $L=0$ to $L = L_0$, with a larger energy than the homogeneous one, and type 1 from $L=L_0$ to $L \rightarrow \infty$ with a smaller energy than the homogeneous one. The situation is similar in the case $n=1$, except the first branch makes a U-turn
, giving two parallel lines at large $L$.
Right panel: Gibbs energy difference $\Delta E$ of the five different types of solutions with one or two solitons 
as functions of $L/L_0$.
Solid lines: 
Types 5 and 7 of Fig.~\ref{fig:9-traj} (which are degenerate when $c_3=c_1$);
dashed lines: 
Types 6 and 8 (also degenerate for $c_1 = c_3$), and dotted lines: 
type 9. These solutions have a larger energy than the homogeneous one when they appear. Their energy is also always larger than that of 
type 1. }\label{fig:ene-9-eq}
\end{figure}

As can also be expected, solutions with either one soliton or two, corresponding to types 5 to 9 in Fig.~\ref{fig:9-traj}, have a larger energy than the other solutions for a given $L$. 
It is therefore unlikely that they play an important role in the time evolution of the system. 
All solutions (except those of type 1 or type 3 with $n=0$) 
extend to $L=\infty$ or not depending on the parameters of the black hole laser. A straightforward calculation shows they actually extend to infinity 
only if the inequality of \eq{eq:inf} is satisfied. 
When it is not, as explained in Appendix \ref{App:single-h}, 
series of solutions terminate at a finite value of $L$ by merging with each other. A series of type 2 solutions will merge with one of type 6 and one of type 4 with one of type 8. The four series types 3, 5, 7 and 9 all merge.  
Instead, series of solutions of type 1 never terminate. This is important because the type 1 with $n=0$ gives the ground state of the system. We now study this case with more details.

\begin{figure}
\begin{center}
\includegraphics[scale=0.9]{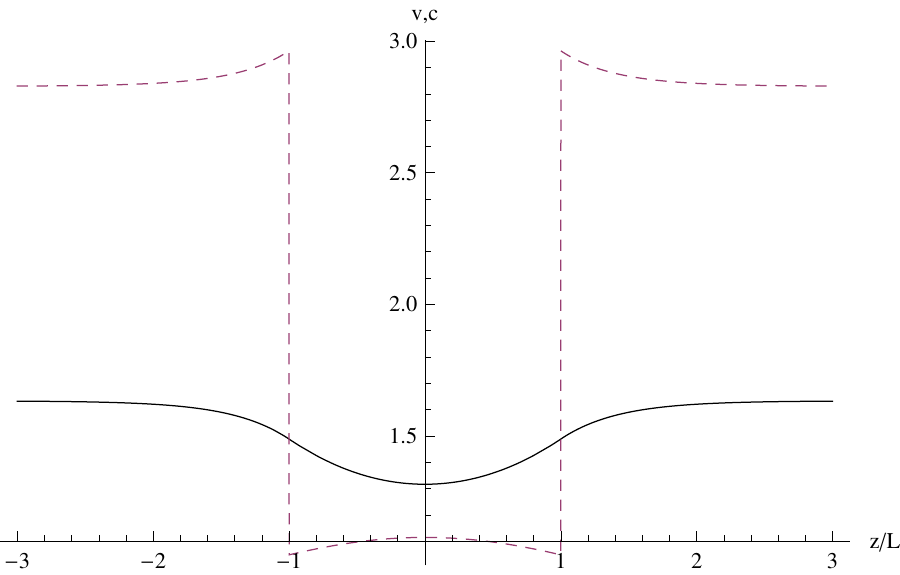}
\includegraphics[scale=0.9]{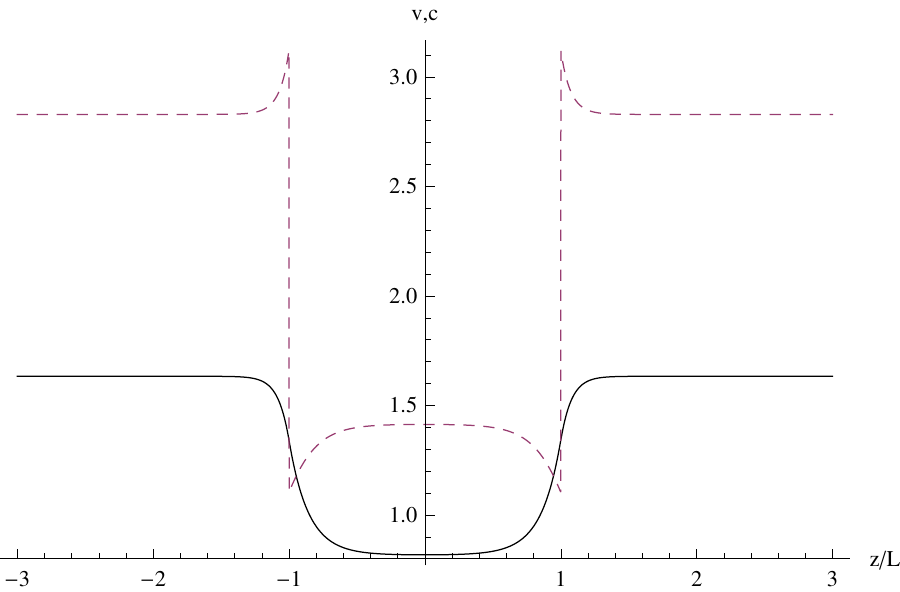}
\caption{Velocity flow (solid line) and sound speed (dashed line) as functions of $z/L$ for the solution with the lowest Gibbs energy $E$ for a distance $L$ slightly above the threshold, $L =1.29\, L_0$ (left diagram), and well above the threshold $L = 7.0 L_0$ (right diagram). The parameters are: $c_1=c_3=2 \sqrt{2}$, $c_2 = 1$, $v=\sqrt{8/3}$ and $f_0=1$. On the right panel, one clearly sees the saturation of the solution with a flat 
profile in the central region which corresponds to a subsonic flow. Notice that the density profile $f^2(z)$ can be deduced from that of $v$ since 
the current $J = f^2(z) v(z)$ is constant. 
}\label{fig:ground-state}
\end{center}
\end{figure}

In this state, for $L \gg L_0$, the amplitude $f$ and velocity $v$ become nearly piecewise constant
with two transition regions at $z \approx \pm L$ of the order of the healing length; see the right panel of Fig.~\ref{fig:ground-state}. In addition, the condensate is subsonic outside the two transition regions. Since negative energy fluctuations only exist when the supersonic flow has a sufficiently large extension, it is clear that this configuration is energetically stable, and represents the end point evolution of the black hole laser effect (if the dynamics leads to stationarity and minimization of the Gibbs energy). 
We now understand that the physical mechanism 
which stabilizes the laser effect is the accumulation of atoms in the central region. Indeed, 
the associated increase of the density reduces the velocity of the flow $v$, and increases the sound speed,
thereby removing the supersonic character of the flow. 
Obtaining the profile of the ground state is the main result of this paper. 
It can be done by using the following procedure. 
The trajectory in phase space $(f,p \equiv f')$ is given by \eq{keyE}, 
with the constants in $I_1$ and $I_3$ given by
\be 
C_i &=& (2 \, v^2 + c_i^2) f_0^2. 
\ee
The third constant $C_2$ is fixed by the value of $L$ through \eq{eqs-num-beg}. 
The profile is then obtained by integration of \eq{keyE}
which is a first-order ordinary differential equation. A convenient 
initial condition is the value of $f$ at $z=- L$, given by $f_{\text{inter},+}$ of \eq{finter}.

So far we have worked with an idealized description where the parameters $g$ and $V$ entering \eq{GPE1} are piecewise constant with two discontinuities. However, in a realistic setup, $g$ and $V$ will change over some finite 
length scale, considered below to be the same and called $\lambda_g$. 
To determine the validity range of results obtained with the steplike approximation, we should determine the leading deviations of our results due to a small $\lambda_g \neq 0$. To this end, we replaced the piecewise constant $g$ and $V$ by various smooth profiles and solved \eq{eq:f} numerically using an imaginary-time evolution. To leading order in $\lambda_g/L_0$, where $L_0$ is given in \eq{eq:Lm}, the only effect is to change the critical values of $L$ where new unstable modes appear.   
Here $L$ is still defined as half the length of the supersonic region. In particular, for the  ground state of
the system, we checked that the relation between the maximum value of $f$ and $L-L_0$, written below in the symmetric
case $c_1^2-v^2 = v^2-c_2^2$ for simplicity,\footnote{\eq{dvtf} can be straightforwardly derived  from \eq{eqs-num-beg} in the case $\lambda_g=0$.}
\be \label{dvtf}
f_{\max }/f_0 -1  = 2 \frac{\left(L-L_0\right) \left(v^2-c_2^2\right)^{3/2}}{\sqrt{2} v^2+\frac{1}{\sqrt{2}}c_2^2} + \mathcal{O}(\left(L-L_0\right)^2),
\ee
is {\it unchanged} to lowest order in $\lambda_g$, although the value of $L_0$ changes. 
We should thus analyze how this value is affected by $\lambda_g \neq 0$. In the general case, when $\lambda_g/L_0 \lesssim 1/10$, we found that the leading deviation of $L_0$ is linear in $\lambda_g$. For profiles which are symmetric between the subsonic and supersonic regions, we found that the differences are quadratic in $\lambda_g$.   
This robustness is in agreement with the spectral analysis of~\cite{2regimesFinazzi} performed in the case of a single horizon. In that case it was found that the Bogoliubov coefficients encoding the scattering across a supersonic transition are well approximated by their steplike approximate values whenever $\lambda_g$, i.e., roughly speaking the inverse of surface gravity, is a tenth of the healing length; see Fig.~4 in~\cite{2regimesFinazzi} for more details. With the observation that \eq{dvtf} remains unchanged at leading order, we have established that the robustness of the step-like approximation extends to the saturation process. 

To end this section we briefly comment on the changes brought about by different sound velocities in $I_1$ and $I_3$. The analysis is very similar to that in the case $c_1 = c_3$ with three phase portraits instead of two. 
The set of solutions is qualitatively similar. In particular, solutions are characterized by the same set of parameters. There is one additional solution for a limited range of $L$ with a larger energy than that of the uniform solution. The other differences are that the first nonuniform solution does not extend to $L=0$ anymore and that previously degenerate solutions now have different energies.

\section{Next-to-quadratic effects and saturation}
\label{thermo}

There exists a close correspondence between the linear analysis of Sec.~\ref{Slt} 
and the nonlinear solutions of Sec.~\ref{stat_sol}. 
Indeed, when $c_1 = c_3$, each degenerate ABM appears at $L = L_m$
for an integer value of $m$, together with a series 
of stationary solutions of type 1 which possess a smaller Gibbs energy than the homogeneous one.
Moreover, for $L < L_m$, the QNM which turns into this ABM when its frequency crosses the real axis corresponds
 to solutions of type 3, with a larger Gibbs energy. In addition, each nondegenerate ABM appears at $L = L_m$ for a half-integer value of $m$ with a series of stationary solutions of types 2 and 4. 

However, this correspondence is not manifest when using the exact treatment  of Sec.~\ref{stat_sol}. 
In this section, we introduce a simplified energy functional $E_s$ which displays  
very clearly the correspondence near $L \approx L_m$ 
for integer values of $m$. For half-integer values of $m$, the analysis is more complicated, 
as briefly explained at the end of the section. 
To construct the functional we use an expansion at lowest nonquadratic order. 
The present perturbative treatment, being rather general, might allow for extensions 
to other cases where zero-frequency waves with large amplitudes are also found, 
for instance, in hydrodynamics~\cite{Coutant_on_Undulations,water_waves},
in massive theories of gravity~\cite{2013arXiv1309.0818B}, and in the 
presence of extra dimensions~\cite{Gauntlett_helical_BHs,Gauntlett_QNM}.
In this construction, we have been inspired by the analysis used in \cite{Baym, Pitaevskii} to describe 
the occurrence of spatially modulated phases in superfluids with a roton-maxon spectrum, when the flow velocity slightly exceeds the Landau velocity. In that case, the effective energy functional which governs the saturation of the amplitude is quartic, as in standard second-order phase transitions. In the present case instead, the stabilizing term 
is cubic, as in a $\lambda \phi^3$ theory. This odd term is due to the breaking of the $\mathbb{Z}_2$ symmetry $f \rightarrow 2 f_0 - f$ discussed in Appendix~\ref{App:M-matrix}. In what follows, we concentrate on even solutions without a soliton, corresponding to types 1 and 3 in Fig.~\ref{fig:9-traj}. For definiteness, we set $c_3 = c_1$.
Then local extrema of a simplified energy functional allow us to recover the change of stability occurring at all $L=L_n$, $n \in \mathbb{N}$. 

\begin{figure}
\begin{center}
\includegraphics[scale=0.7]{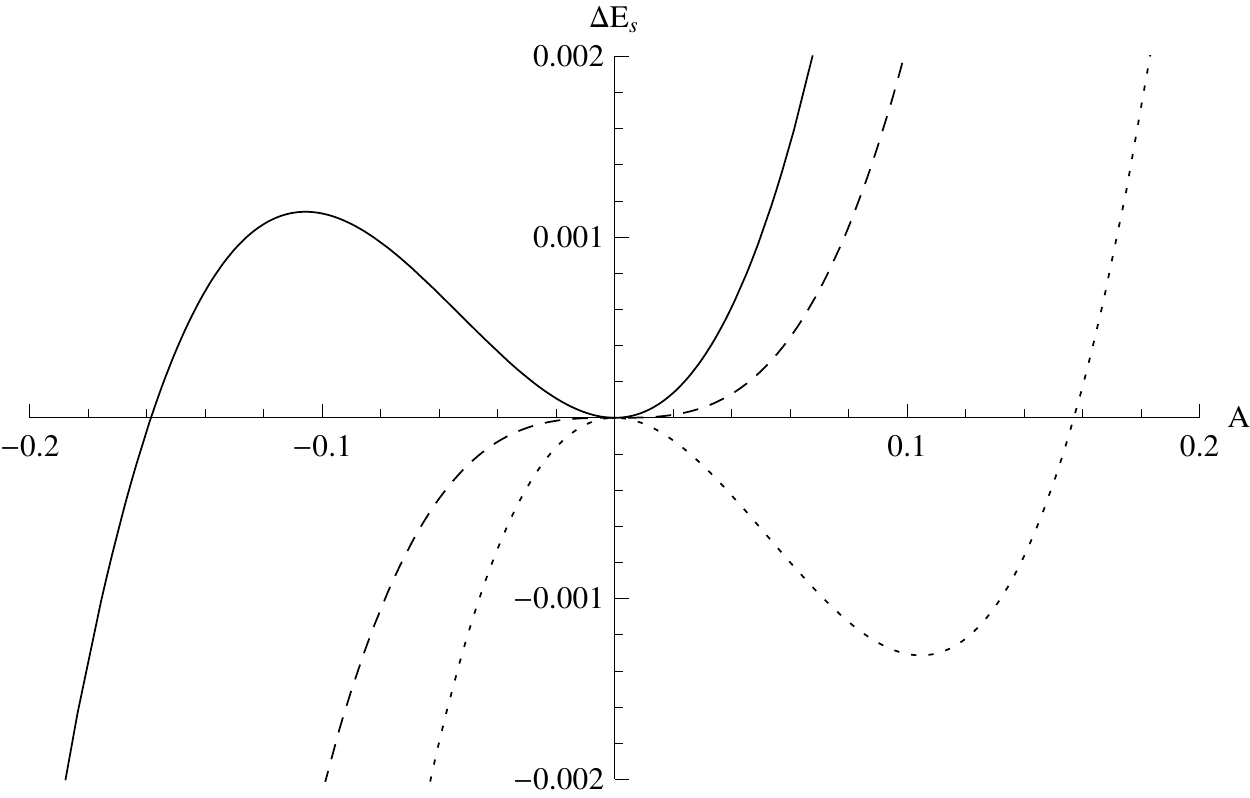}
\includegraphics[scale=0.55]{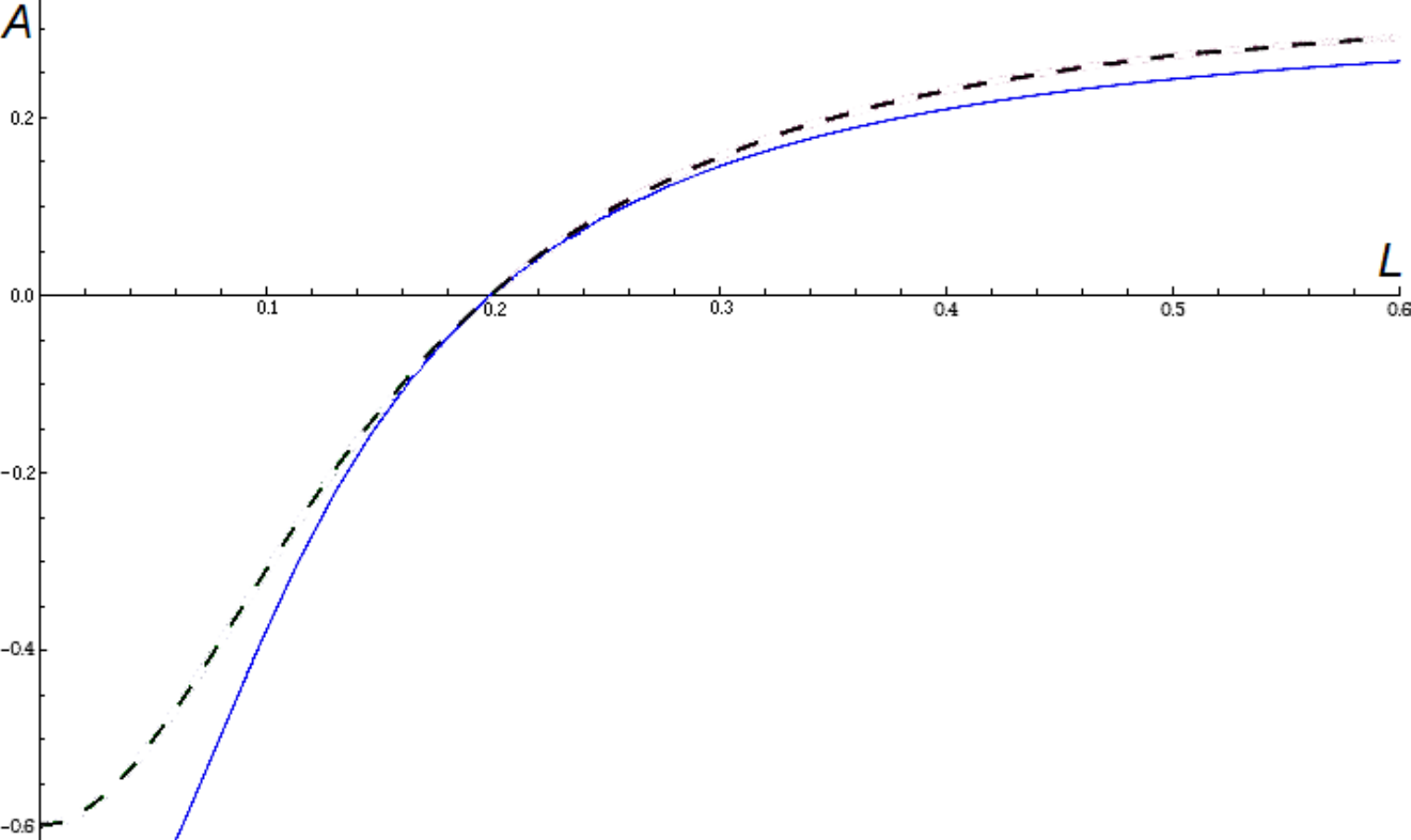}
\caption{Left panel: Simplified Gibbs energy (\ref{eq:en_A}) 
as a function of the amplitude $A$ for $L=0.85 L_0$ (solid line), $L= L_0$ (dashed line), and $L=1.2 L_0$ (dotted line). The wave vector is $k_2=2 \sqrt{v^2 - c_2^2}$; seen
\eq{eq:lambda}. One clearly sees that the change of stability of the homogeneous solution
occurs for $L= L_0$. Right panel: Amplitude $A$ of the inhomogeneous solution (type 3 for $A<0$ and type 1 for $A>0$) as a function of $L$. The solid line is the result from the simplified treatment of (\ref{eq:AB}), and the dashed line is from the full nonlinear solution. The two methods give the same values for $A$ and $\pd_L A$ at $L=L_0$. The parameters are: $c_1=2.0$, $c_2=0.5$, $v=1.0$, $f_0=1.0$.} \label{fig:EBaym}
\end{center}
\end{figure}

As in Appendix~\ref{App:M-matrix}, 
we write $f(z)=f_0 + \delta f (z)$, where $f_0$ is the globally homogeneous solution. To third order in $\delta f$, the Gibbs functional $E$ reads (up to a constant term)
\be \label{eq:third_order}
\Delta E=\int_{-\infty}^{\infty}  \left(\frac{1}{2}\left(\frac{\partial \delta f}{\partial z}\right)^2+2\left(c(z)^2-v^2\right)\delta f^2\right) dz+2\int_{-\infty}^{\infty}   \left(c(z)^2+v^2\right) \frac{\delta f^3}{f_0} \, dz+\text{...}
\ee
The idea is now to choose an ansatz for $\delta f$ which depends on
some parameters, and extremize $E$ with respect to them.  
If the ansatz is well chosen, the solution will be close to the exact solution.
To optimize the choice near $L_n$, we work with an ansatz compatible with the linear even solutions of \eq{linearfth}
\be \label{ansatz}
\delta f(z)=
 \left\lbrace  
\begin{array}{ll}
 A_1e^{-k_1|z|} , & \text{for} |z|>L , \\
 A \cos \left(k_2 z\right),  & \text{for} |z|<L .
\end{array}
\right.
\ee
Continuity and differentiability at $|z|=L$ give
\be 
k_1=k_2 \tan \left(k_2 L\right) , 
\ee
and 
\be 
A_1=A\cos \left(k_2 L\right)e^{k_2 L \tan \left(k_2L\right)}.
\ee 
Performing the integrals explicitly, the simplified version of \eq{eq:third_order} becomes
\be \label{eq:en_A}
\Delta E_s = W_2(k_2,L)\,  A^2 + W_3(k_2,L) \, A^3+ \mathcal{O}(A^4),
\ee
where 
\be \label{eq:W2}
W_2=2\left(\left(\frac{k_2^2}{4}+c_2^2-v^2\right) L +\left(c_2^2-c_1^2\right)\frac{\sin \left(2 k_2 L\right)}{2 k_2}+\left(c_1^2-v^2\right)\frac{\cos \left(k_2 L\right) }{k_2 \sin \left(k_2 L\right)}\right) ,
\ee
and
\be \label{eq:W3}
W_3= \frac{4}{f_0} \lp\frac{c_2^2+v^2}{k_2}\sin \left(k_2 L\right)\left(1-\frac{\sin \left(k_2 L\right)^2}{3}\right)+ \frac{c_1^2+v^2}{3 k_2} \frac{\cos \left(k_2 L\right)^4}{\sin \left(k_2 L\right)}\rp . 
\ee
Because of the term of order 3, the simplified Gibbs energy (\ref{eq:en_A}) seen as a function of $A$ at fixed $k_2$ is not bounded from below (see Fig.~\ref{fig:EBaym}). Adding higher-order terms would not solve this issue. A straightforward calculation shows that in spite of the positive contribution from $\frac{1}{2} g f^4$, the quartic term is always negative. In addition, all higher even-order terms have negative coefficients because they are all obtained from the expansion of $-J^2/(2 f^2)$. This does not signal an instability of the system, rather it limits the validity of the ansatz of \eq{ansatz}. 

The extremization proceeds in two steps. First we extremize \eq{eq:en_A} with respect to the amplitude.
Then the optimal value of $k_2$ is found by extremizing the result with respect to $k_2$.
We start by examining the situation for $L$ near $L_0$. 
At fixed $k_2$, Eq. (\ref{eq:en_A}) has two extrema (see Fig.~\ref{fig:EBaym}, left panel): 
a local minimum and a local maximum, which can be interpreted as a metastable and an unstable solution respectively. 
One extremum corresponds to 
$A=0$, \textit{i.e.} to the homogeneous solution. It is metastable if $W_2>0$ and unstable if $W_2<0$. The other extremum describes 
an inhomogeneous solution. 
Its amplitude is 
\be \label{eq:AB}
A=-\frac{2 W_2}{3 W_3} , 
\ee
and its Gibbs energy is 
\be \label{GEA}
\Delta E_s^{\rm inhom} = \frac{4 W_2^3}{27 W_3^2}.
\ee 
On the right panel of Fig.~\ref{fig:EBaym}, we compare 
\eq{eq:AB} for $k_2 =2 \sqrt{v^2 - c_2^2}$
with the exact value of the amplitude, defined as $f(z=0)-f_0$. 
Near $L = L_0$ we have a very good agreement between the two which demonstrates 
that \eq{eq:en_A} correctly describes the relevant field configurations involved 
in the destabilization of the homogeneous solution.
This agreement is guaranteed by the 
facts that, to quadratic order, our ansatz \eq{ansatz} is exact  
and that the third-order term does not vanish. Indeed, it is easily shown that terms coming from a more accurate ansatz would be at least fourth order in the amplitude. 

\begin{figure}
\begin{center}
\includegraphics[scale=0.7]{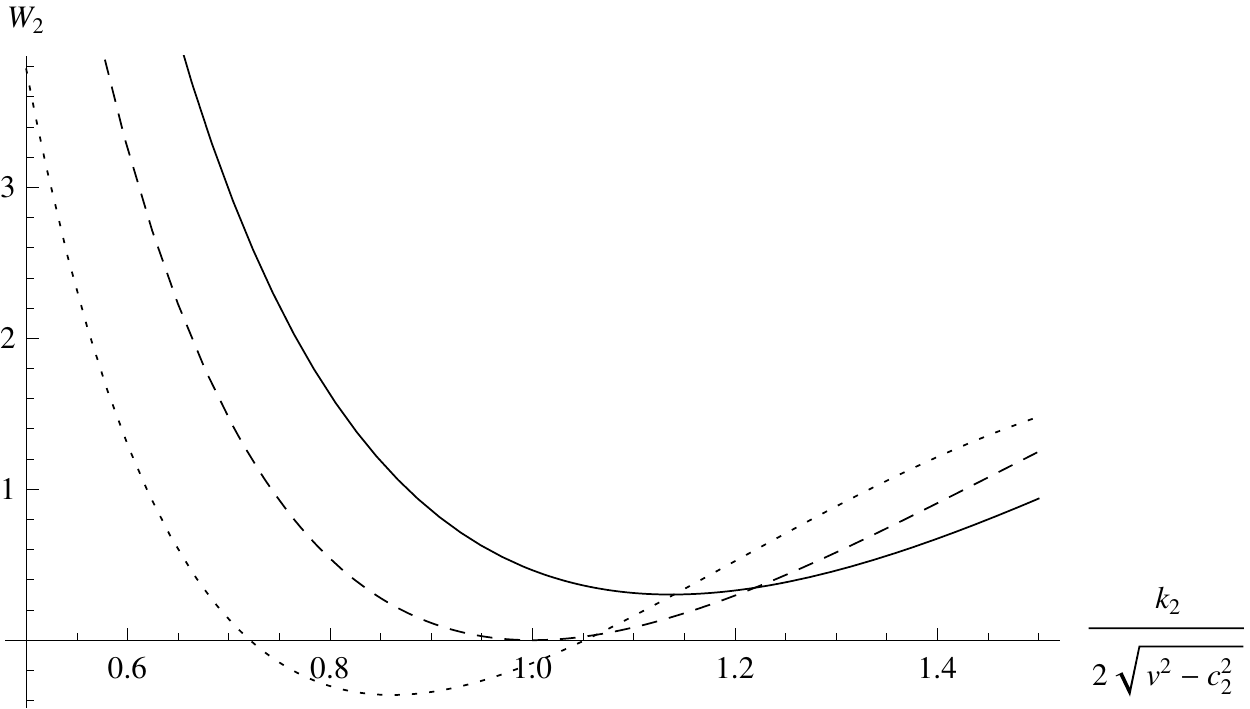}
\includegraphics[scale=0.9]{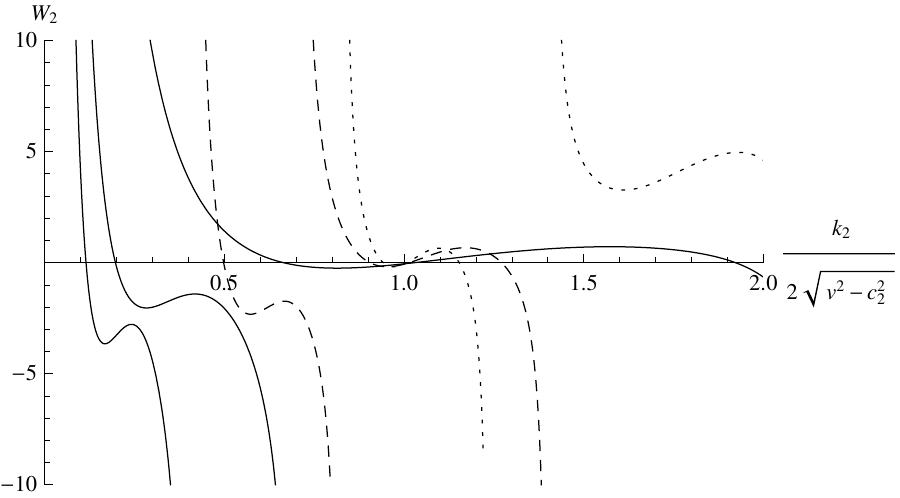}
\caption{Left panel: Coefficient $W_2$ of the quadratic term in the Gibbs energy $E$ as a function of $k_2/(2 \sqrt{v^2 - c_2^2})$ for $L=0.85 L_0$ (solid line), $L= L_0$ (dashed line), and $L=1.2 L_0$ (dotted line). 
We see that the instability occurs for $L = L_0$, and that 
the wave vector of the unstable mode is exactly  $k_2 = 2 \sqrt{v^2 - c_2^2}$. Right panel: The value of $W_2$ as a function of $k_2/(2 \sqrt{v^2-c_2^2})$ along the branches $n=0$ (solid line), $n= 1$ (dashed line), and $n=2$ (dotted line) for $L=1.2 L_0$ , $L=1.2 L_0+\lambda_0/4$ and $1.2 L_0+\lambda_0/2$. We see that $W_2$ has an infinite series of local minima for $k_2 \to \infty$. 
They describe solutions of type 3 when the local minimum is positive, and type 1 when it is negative.
As $L$ increases these minima migrate to lower values of $E$ and of $k$.
For $L_n < L < L_{n+1}$, $n+1$ minima have a lower energy than the homogeneous solution.
 The parameters of both panels are: $c_1=2.0$, $c_2=0.5$, $v=1.0$, $f_0=1.0$.} \label{fig:w2}
\end{center}
\end{figure}

It is also interesting to study the dependence of $W_2$ in $k_2$. In Fig.~\ref{fig:w2},
this is represented for three values of $L$, slightly below, at, and above $L_0$. 
For $L<L_0$, one sees that $W_2$ remains positive for all values of $k_2$, 
which confirms that the homogeneous solution is stable for all these perturbations.
We also see that the first mode which becomes 
unstable corresponds to $k_2 = 2 \sqrt{v^2-c_2^2}$, in agreement with 
\eq{eq:lambda}. The sign change of  $W_2$ at $L = L_0$ precisely
corresponds to the transition from 
type 3 for $L< L_0$ to 
type 1 for $L> L_0$. 

It is rather easy to consider the other sectors with $n > 0$. As the right panel of Fig.~\ref{fig:w2} shows, $W_2$ has an infinite set of local minima in $k_2$. 
The minima increase with $n$ 
and decrease with $L$ following $W_{2,n}^{min} \sim \frac{\pi^2 n^2}{4 L}$. 
For any positive integer $n$, the $n$th minimum becomes negative for some value 
of $L$, which is given by $L_n$ of \eq{eq:Lm}.
Notice that the corresponding value of $k_2$ is always $2 \sqrt{v^2 - c_2^2}$ irrespective
of the value of $n$. This signals the birth of a new instability of the homogeneous solution as well as 
the beginning of a new series of metastable nonlinear solutions. 
This can be understood from the behavior of \eq{eq:W2} and \eq{eq:W3} under a change of $n$. 
Indeed, when  $k_2 = 2 \sqrt{v^2 - c_2^2}$, the first term in \eq{eq:W2} vanishes.
As a result, $W_2$ is unchanged under $L \rightarrow L +\lambda_0/2$, while $W_3$ and $\pd_{k_2} W_3$ 
flip signs.
This simply reflects that adding one wavelength to the solution in $I_2$
replaces a minimum at $z = 0$ by a maximum. A straightforward calculation shows that $\pd_{k_2} W_2$ is also invariant.
So, for all $n \in \mathbb{N}$, $k_2 = 2 \sqrt{v^2-c_2^2}$ remains the value of $k_2$ where 
a change of stability occurs for $L = L_{n}$, as was the case for $L=L_0$.

It is also possible to use the initial velocity $v$ as a control parameter instead of $L$. The analysis is then very similar. If $v = c_2$ there is no unstable mode, which translates as the absence of a negative local minimum in $W_2(k_2)$. This is because $L_0$ (as well as $\lambda_0$) is infinite, so that any finite $L$ is smaller than $L_0$. When $v$ is increased from $c_2$ to $c_1$, $L_0$ decreases monotonically from $\infty$ to $0$. The first unstable mode appears when $L_0$ becomes equal to $L$. Then other unstable modes arise each time 
$L_0 + n \lambda_0/2 = L$ for some integer $n$. $\lambda_0$ is also monotonically decreasing in $v$ but remains finite in the limit $v \rightarrow c_1$, with a limiting value given by
\be 
\lambda_{0,min} = \frac{\pi}{\sqrt{c_1^2 - c_2^2}}.
\ee 
The number of stable or metastable inhomogeneous solutions at fixed $L$ thus goes 
from $0$ for $v=c_2$ to 
\be 
\left\lfloor \frac{4 L}{\pi} \sqrt{c_1^2 - c_2^2} \right\rfloor +1 
\ee
for $v=c_1$.

So far we have discussed the transition occurring for integer values of $m$.
The stability changes associated with $L \approx L_m$ with a half-integer $m$ are more subtle 
for the following reasons.
When $c_1 \neq c_3$, one series of solutions (of type 2 if $c_1 > c_3$ or type 4 if $c_1 < c_3$) extends up to $L = L_m' < L_m$ with a larger Gibbs energy than the homogeneous solution. The other one (type 4 if $c_1 > c_3$ or type 2 if $c_1 < c_3$) then exists only for $L > L_m$ with a smaller Gibbs energy. When $c_3 \rightarrow c_1$, $L_m' \rightarrow L_m$ and the two series of solutions become degenerate. This change of behavior has important consequences for the analysis presented above. If $c_1 \neq c_3$, it is still the third-order term which governs the saturation and the expansion of the energy functional accurately describes the change of stability. However, if $c_1 = c_3$ the third-order term vanishes. One must then include contributions which are of order 4 in the amplitude and choose a more accurate ansatz than that provided by linearized solutions. This makes the analysis technically more involved and hides the intrinsic simplicity of the procedure. Similarly, the study of types 5 to 9 requires expanding the Gibbs energy functional around a solution with one or two solitons, leading to computational difficulties. 

\section{Conclusions}

To perform both a linear and a nonlinear stability analysis,
we used a simple model of black hole lasers in one-dimensional infinite Bose-Einstein condensates. 
The simplicity is due to the use of a piecewise constant potential which is such that there 
exists an exact solution with a uniform flow velocity, while the sound velocity has two discontinuities. 
Using the linearized mode equation and matching conditions, 
the set of complex frequency modes that are responsible for the dynamical instability has 
been explicitly obtained. In particular we showed that each new unstable
mode arises in two steps. For a finite interval of 
the distance $2L$ between the two discontinuities, we found that the unstable mode has a purely imaginary frequency. For larger values we recovered the situations found
in~\cite{CoutantRP,FinazziParentani}; seen Fig.~\ref{fig:QNM}. We claim that this 
two-step 
process will also apply to smooth profiles, at least when the gradients of the potential $V$, and the coupling $g$, are sufficiently large in the units of the inverse healing length. 
Indeed, in this limit, on the first hand, it has been shown~\cite{MacherBEC,2regimesFinazzi} that the Bogoliubov coefficients encoding the mode mixing at each sonic horizon are in close agreement with those derived from the matching conditions we used. Hence the solutions of $\det M = 0$ should continuously depend on the gradients. 
On the other hand, we found that the dimensionality of the unstable sector is 1 when the frequency is purely imaginary, 
and not 2 as is the case when the frequency is complex. Therefore, it will remain 1 even if the value of the imaginary frequency is slightly shifted,
and these frequencies will remain purely imaginary. 

To find the end point of the evolution of this dynamical instability, we characterized the stationary nonlinear solutions of the GPE with a finite Gibbs energy. We showed that a set of nine nonlinear solutions corresponds to each unstable mode, and we explained the origin of this multiplicity; seen Fig.~\ref{fig:9-traj}. We also showed that in each set, one solution can be conceived as the end-point evolution (in the mean field approximation since we work with solutions of the GPE) of the corresponding instability; seen Fig.~\ref{fig:ene-9-eq}. When considering the whole set of solutions at fixed $L$ we identified the lowest energy state and studied its properties. In particular we numerically verified that the maximum value of the density is, at leading order, unchanged when replacing our discontinuous profiles by continuous ones which are sufficiently steep. In the steplike regime, we analytically constructed the exact solutions by pasting building blocks consisting 
of 
exact solutions of the GPE associated with each 
homogeneous region, 
see Appendix~\ref{App:single-h}. To explicitly relate the onset of instability described by the complex frequency modes of Sec.~\ref{Slt} to the nonlinear solutions of Sec.~\ref{stat_sol}, 
we presented in Sec.~\ref{thermo} a treatment based on a Taylor expansion of the energy functional and a simplified ansatz which displays the second-order transition between the homogeneous solution and a spatially structured one. 

A natural extension of this work would be to investigate the time evolution of this system, from the initial instability to the final configuration. To identify the validity domain of our findings, it would also be interesting to work beyond the mean field approximation, and to consider in more detail smooth profiles in which the initial sound velocity is continuous. Finally, computing the spectrum on top of the various stationary solutions would allow for a more precise stability analysis and tell us whether there can be long-lived metastable states.   

\smallskip \smallskip 
\smallskip 
{\it Note added.—}

\noindent
We would like to mention that the transition from the initial unstable homogeneous solution to the lowest-energy state described in Sec.~\ref{stat_sol} provides an interesting example of a process which mimics a unitary black hole evaporation. 
When considering gravitational black holes, we remind the reader that it is still unknown whether the evaporation process is nonunitary, as originally suggested by Hawking, or if it satisfies unitarity, as is the case for standard quantum mechanical processes. We also remind the reader that in order for the emitted Hawking radiation to end up in a pure state at the end of the evaporation (when starting from a pure state), it is necessary to have a nondegenerate final black hole state.
As argued by Page~\cite{Page}, this implies that the Hawking quanta emitted after a certain time must be correlated to the former ones. Using a mean field treatment of the metric, that is, when adopting the so-called semiclassical scenario,
this conclusion is highly nontrivial since the Hawking quanta are entangled with their negative energy partners~\cite{Primer,MassarRPFromPRD1995} but are uncorrelated with each other.
One can of course hope that when working beyond the mean field approximation, quantum backreaction effects will restore the unitarity. The difficulty one then faces is to find some microscopic description of black holes in which this can be shown to occur. The main virtue of the present model is that it combines in a nontrivial way two essential elements. First, at early times, using a linearized treatment, the emitted phonons can be shown to be entangled with the negative energy partners which are trapped in central region $I_2$, as is the case for the Hawking process.
~\footnote{After a while, as noticed in~\cite{BHL}, because the laser effect is taking place, there exist correlations among the emitted quanta. However these correlations are not sufficient to restore unitarity, as the correlations to the partners are still present.} 
Second, the full  Hamiltonian possesses a unique ground state. One can therefore deduce, like Page, that after some time, the emitted phonons will be correlated with the former ones. Another virtue of the model is that these correlations should, in principle, be calculable without encountering the uncontrolled divergences which occur in perturbative treatments of quantum gravity. We hope to study these questions in the near future.

\acknowledgements 
We are grateful to Iacopo Carusotto, Anatoly Kamchatnov, Ted Jacobson, Nicolas Pavloff, Gora Shlyapnikov 
and Robin Zegers for advice and interesting conversations. We are also thankful to Antonin Coutant and Stefano Finazzi for comments on an early version of this work.
F.M. acknowledges financial support from the \'Ecole Normale Supérieure of Paris. 

\appendix

\section{Structure of the equation on complex frequencies}
\label{App:M-matrix}

In this appendix we detail the procedure we used to find the ABM and QNM. 
The explicit form of the matrix $M$ whose determinant encodes the matching conditions is shown and the results are compared with the Bohr-Sommerfeld approximation used in \cite{FinazziParentani}.

\subsection{Complex frequency modes}
The procedure to find ABM and QNM consists of two steps. First we solve the linearized Gross-Pitaevskii equation (GPE) in each of the three regions $I_1$, $I_2$ and $I_3$ and impose boundary conditions at $z \rightarrow \pm \infty$ to retain the solutions which are "outgoing" in a generalized sense, which we will explain.
Then we impose matching conditions at the two horizons $z = \pm L$ to find the globally defined modes. 

\begin{figure}
\begin{center}
\includegraphics[scale=0.5]{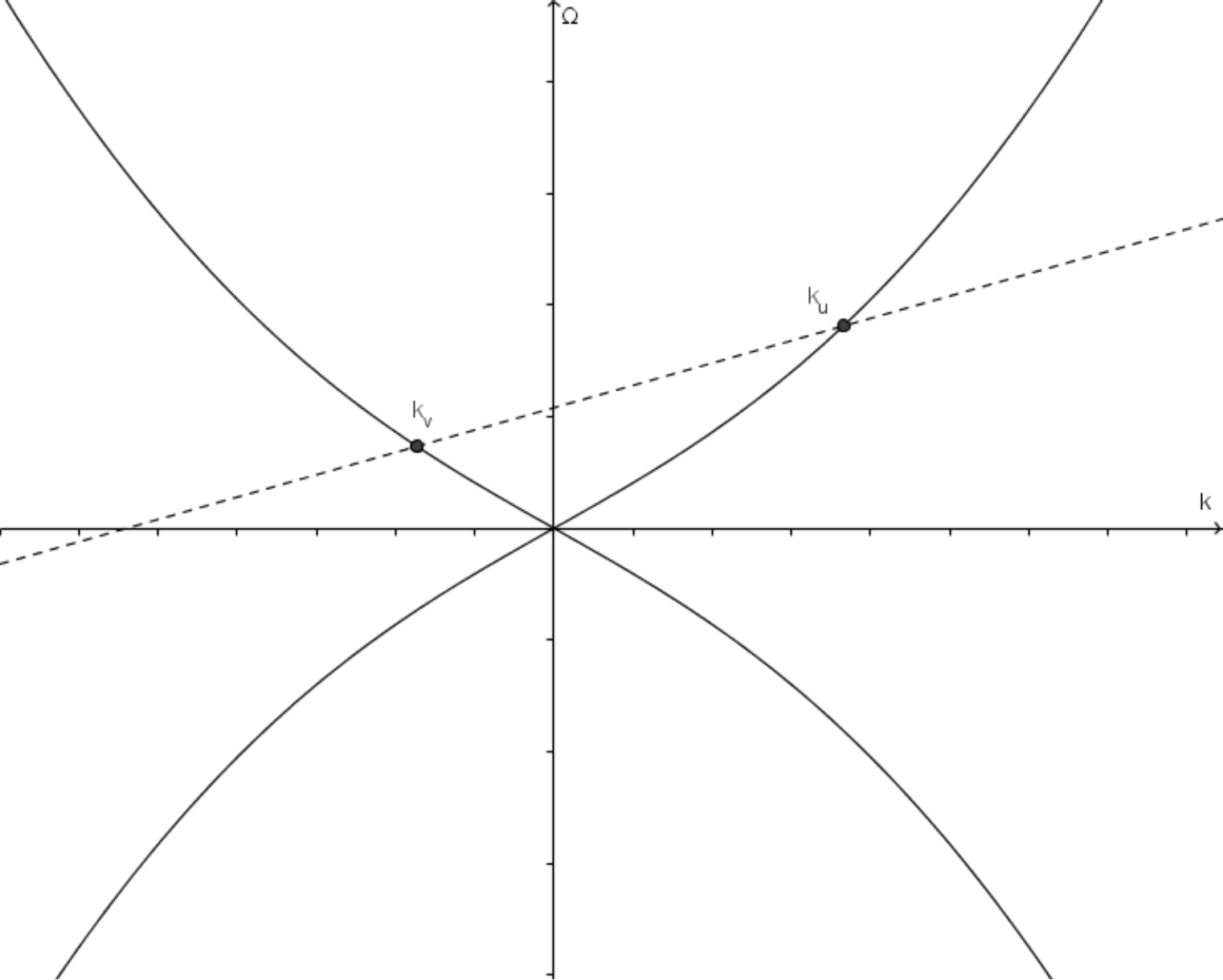}
\includegraphics[scale=0.5]{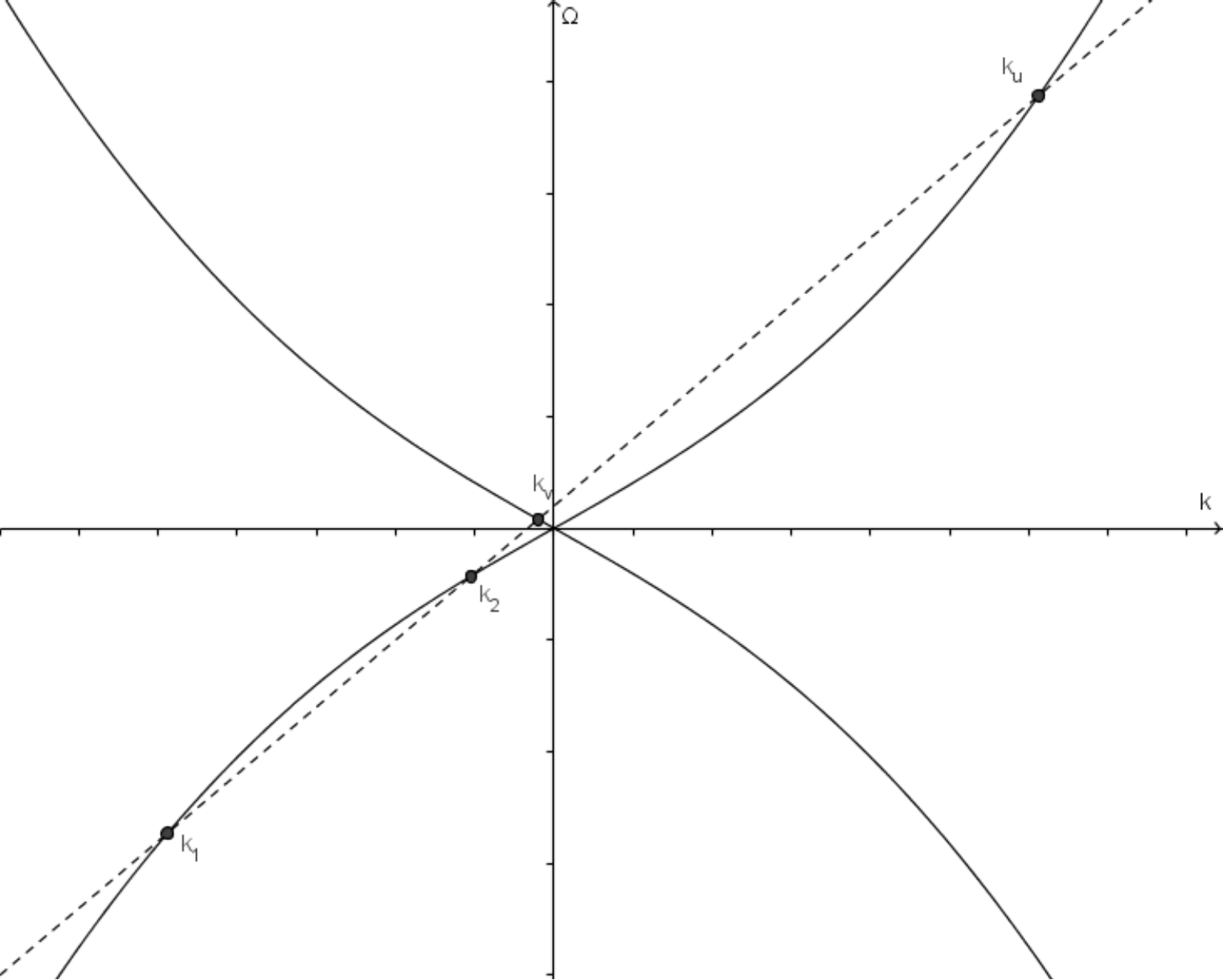}
\caption{Graphical resolution of the dispersion relation in a subsonic (left diagram) or supersonic (right diagram) flow. The solid curve represents $\Omega(k)$ of \eq{eq:disprel1} and the dashed line $\omega - v k$. Here $\om$ is real and positive. In the subsonic case there are two real roots: a left mover $k_v$ and a right mover $k_u$. In the supersonic case and if $\om$ is small enough there are two additional real roots with $\Omega < 0$: $k_1$ and $k_2$. $k_1$ is a right mover while $k_2$ is a left mover.}\label{fig:disprel1}
\end{center}
\end{figure}

\begin{figure}
\begin{center}
\includegraphics[scale=1.0]{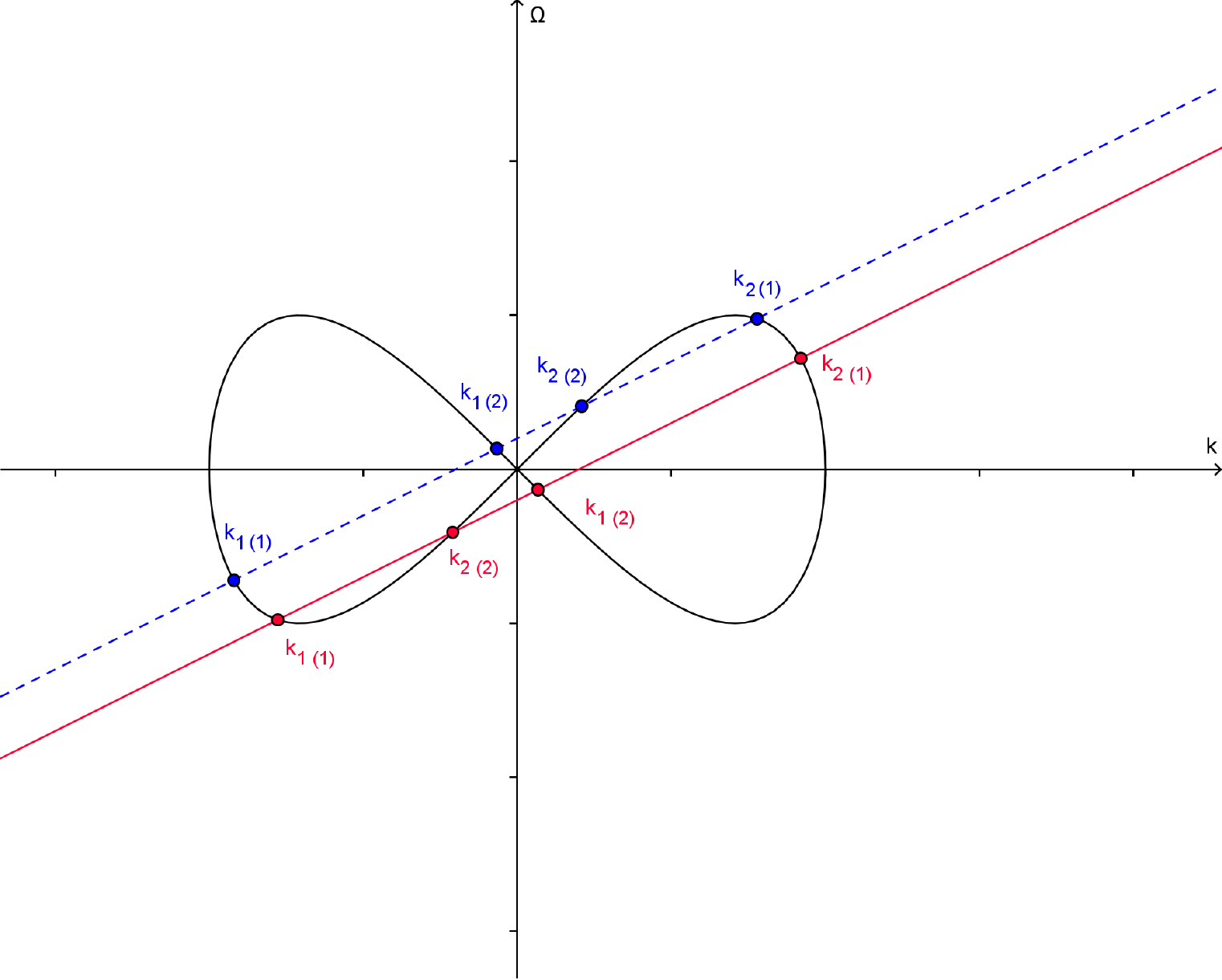}
\caption{Graphical resolution of the dispersion relation for $\om \in i \mathbb{R}$. Solid curve: $i \Omega$ as a function of $i k$ for $k \in i \mathbb{R}$. Dashed blue line: $i (\omega - v k)$ as a function of $i k$ for a subsonic flow and $0<-i \om<\Gamma_0$, where $\Gamma_0$ is the positive value of $-i \om$ at which two roots merge.
In that case there are four purely imaginary roots to the dispersion relation. We set $c_3 = c_1$. $k_j^{(1)}$ and $k_j^{(2)}$ are the two modes we use in region $j$ to build ABM. Solid red line: $i (\omega - v k)$ as a function of $i k$ for a subsonic flow and $-\Gamma_0<-i \om<0$.}
\label{fig:disprel_im_sub}
\end{center}
\end{figure}

We write $\psi(t,z)= (f_0 + \delta f(t,z))\, e^{i (\theta_0(z)+\delta \theta(t,z))}$, where $f_0 \, e^{i \theta_0(z)}$ is the solution of \eq{GPE1} 
with a uniform amplitude. To first order in $(\delta f, \delta \theta)$, \eq{GPE1} gives
\be \label{linearfth}
\left\lbrace 
\begin{array}{ll}
 \partial_t \, \delta f + v \, \partial_z \delta f + \frac{1}{2} \, f_0 \, \partial_z^2 \delta \theta= 0 &   \\
  - \frac{1}{2} \, \partial_z^2 \delta f +2 \, c_j^2 \, \delta f \,+ \,f_0 \, v \, \partial_z \delta \theta \, + \, f_0 \, \partial_t \delta \theta=0 &  
\end{array}
\right. .
\ee
The solutions given in \eq{eq:fandtheta} determine the dispersion relation of \eq{eq:disprel1}. 
For a given $\om$, there are four solutions. In a subsonic flow, for $|c|>|v|$ and $\om$ real, 
$k_u$ describes the right mover, and $k_v$ the left mover; seen the left panel of 
Fig.~\ref{fig:disprel1}. The two other roots are complex: $k_+$ gives the exponentially decreasing mode at $z \rightarrow + \infty$, whereas $k_-$ is the decreasing one at $z \rightarrow - \infty$.  In the central supersonic region, as can be seen from the right panel of
Fig.~\ref{fig:disprel1}, the four roots are real if $-\om_{max}<\om<\om_{max}$, where
\be \label{eq:omMax}
\om_{max}=2 \sqrt{2} \sqrt{|v| +\sqrt{ v^2 +8 \, c_2^2}} \lp \frac{v^2-c_2^2}{3 \, |v| +\sqrt{ v^2 +8 \, c_2^2}} \rp^{3/2} .
\ee

When looking for ABM, one must keep only the wave vectors with a negative imaginary part in $I_1$,
and a positive imaginary part in $I_3$. When considering the ABM which grows in time, i.e. for $\Im \om = \Gamma > 0$, 
in $I_1$, the two wave vectors respectively correspond to the analytical continuations
of the left-moving mode $k_v$ and the evanescent mode $k_-$.  In $I_3$ instead, they correspond to the right-moving mode $k_u$ and the evanescent mode 
$k_+$; seen Fig.~\ref{fig:disprel1} for $\om \in \mathbb{R}$. 
The modes selected in this way are outgoing in that the analytical continuation of the roots $k_\om$
which are real for real $\om$ possess an outgoing group velocity.
Notice that this definition also applies to the degenerate case characterized by
a purely imaginary $\om$. Indeed, as long as $\vert \Im \om \vert < \Gamma_0$, where $\Gamma_0$   
is given by 
\be\label{eq:Gamma0} 
\Gamma_0=\sqrt{8 \frac{ |v| +\sqrt{ v^2 +8 \, c_1^2}}{\left(3 \,  |v| +\sqrt{ v^2 +8 \, c_1^2}\right)^3}\left(c_1^2-v^2\right)^3} ,
\ee
the various roots do not cross each other; seen Fig. \ref{fig:disprel_im_sub}. Hence, in that interval, 
the complex roots can be viewed as analytical extension of their ancestors defined at $\om = 0$. 

We {\it define} the QNM by the same outgoing condition, but this time
in the complex lower half-plane $\Gamma<0$. We thus also retain $k_v$ and $k_-$ in $I_1$, and $k_u$ and $k_+$ in $I_3$.\footnote{N.B. These conditions differ from those used in Ref.~\cite{Garay}.} 
It is therefore not so surprising that all ABM appear as some QNM cross the real axis. 
Yet, there exists additional QNM which are not the analytical continuation of ABM. 
It would be nice to identify under which conditions our definition of QNM is recovered when analyzing the
poles of the retarded Green function. We hope to answer this question in the near future.

It should be noticed that the matrix $M$ defined below 
possesses a smooth limit $\Re \om \rightarrow 0$. So, the procedure to find purely imaginary frequencies does not differ from the general case. Yet, in this case, the instability is described by a real degree of freedom, instead of a complex one as in the case $\Re \om \neq 0$. This reduction can be seen by considering 
the solutions of the Bogoliubov-de Gennes equation~\cite{Goralectures,Pita-Stringbook}.
Whether $\Re k = 0$ or not, when $\Re \om = 0$, the complex frequency modes
obey $\phi_k=\phi_{-k^*}$. 
The number of degrees of freedom is thus halved with respect to the case $\Re \om \neq 0$. However, the number of matching conditions is also halved, which explains why $det M = 0$ also gives the modes with purely imaginary frequencies. 

\subsection{Structure of the matching matrix $M$}

Continuity and differentiability of $\delta f$ and $\delta \theta$ at the two horizons give eight matching conditions which can be written as eight linear relations between the coefficients of the modes for a given $\om$. A nontrivial solution exists if and only if the determinant of the 8-by-8 matrix $M$ defined below vanishes. 

Lines of $M$ correspond to each of the eight matching conditions, while its columns correspond to the eight modes: the two modes in $I_1$ in the first two columns, the four modes in $I_2$ in the next four columns and the modes in $I_3$ in the last ones. 
The coefficients of the first line of $M$ are the values of $e^{i k z}$ evaluated at $z=-L$ for the corresponding mode, multiplied by $k^2$. The same factor $k^2$ multiplies all the coefficients of a given column, so it does not change the equation $det M = 0$. It is introduced to avoid important numerical errors when $k$ is close to zero. The last two coefficients of the first line are set to zero because the modes in $I_3$ do not contribute at $z=-L$
The second line of $M$ contains the derivative  of $e^{i k z}$ evaluated at $z=-L$, multiplied by $k^2$. 
As for the first line, the last two coefficients are set to zero. 
The third and fourth lines are built the same way, except $e^{i k z}$ is replaced by    $\frac{\delta \Theta}{\delta F} \, e^{i k z}$. So, the first four lines encode the matching conditions at $z=-L$.
The last four lines are constructed similarly, except $-L$ is replaced by $L$ and the first two coefficients are set to zero instead of the last two, since the relevant regions are then $I_2$ and $I_3$. 

Explicitly, the first two columns of $M$ have the form
\be 
\left(
\begin{array}{c}
 k_1^2e^{-i k_1L} \\
 k_1^3e^{-i k_1L} \\
 \Omega_1 e^{-i k_1L} \\
 k_1 \Omega_1 e^{-i k_1L} \\
 0 \\
 0 \\
 0 \\
 0
\end{array}
\right),
\ee 
where $k_1$ is the wave vector of either one of the two modes in $I_1$: $k \in \left\lbrace k_1^{(1)},k_1^{(2)}\right\rbrace$, and $\Omega_1 = \om - v k_1$. 
The seventh and eighth columns have the same structure, with the first four lines replaced by the last four, evaluated with the appropriate roots. The four central columns have no zero, and contain twice the above structure of four entries.

\subsection{Comparison with the Bohr-Sommerfeld approach of \cite{FinazziParentani}}

\begin{figure}
\begin{center}
\includegraphics[scale=0.7]{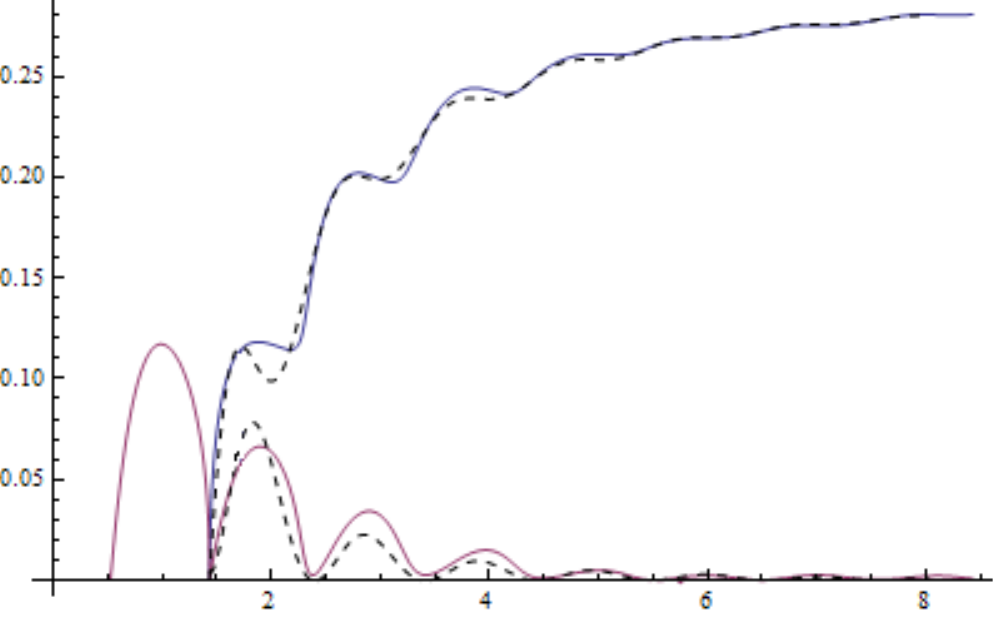}
\caption{Frequencies of the ABM with $n=0$ obtained using the Bohr-Sommerfeld approach,
and by solving $det(M)=0$ as a function of $L$. The solid lines are the results from $det(M)=0$ (blue: real part; purple: imaginary part) and the dashed lines are from the Bohr-Sommerfeld approximation. The parameters are: $c_1=c_3=1.5$, $c_2=0.5$, $v=1.0$ and $f_0=1$.}\label{fig:BS}
\end{center}
\end{figure}

In~\cite{FinazziParentani} a semiclassical (Bohr-Sommerfeld) 
approach was used to study the set of ABM. As usual in this kind of treatment, the set of single-valued solutions is characterized by an integer $n_{\rm BS} = 0, 1, 2, ...$ which gives the integrated phase shift when making a round trip between the two horizons. This approach was shown to be in good agreement with the numerical data when $L$ is large enough for a fixed $n$, as expected because corrections to the semiclassical approximation decrease in this limit. In fact, our exact treatment agrees both qualitatively and quantitatively with~\cite{FinazziParentani} in this limit; seen Fig.~\ref{fig:BS}. In particular, one verifies that the parameter $n_{\rm BS}$ plays exactly the role of $n$ defined in Sec.~\ref{Slt}. There is thus a one-to-one correspondence between the set of ABM found using the two methods. However, while it correctly predicts that new complex frequency ABM appear at $L_{n+1/2}$, the Bohr-Sommerfeld approach completely misses the existence of the degenerate ABM with imaginary frequency which exists for each $n$ for $L_{n}<L<L_{n+1/2}$. This is not too surprising since corrections to the semiclassical approximation are large in this case.

\section{Stationary solutions in the presence of one single horizon}
\label{App:single-h}

In this appendix we discuss the stationary solutions of the GPE in the presence of a single discontinuity, in a black- or white-hole configuration. We focus on solutions for which $f$ goes to a constant $f_0$ at infinity in the subsonic region. These solutions serve as building blocks for the black hole laser solutions. 

We consider a one-dimensional, infinite Bose-Einstein condensate whose two-body coupling $g$ and external potential $V$ are piecewise constant with a single discontinuity at $z=0$. We assume repulsive interactions: $g>0$. The current $J$ is uniform for consistency with the continuity equation. We denote as $g_1$, $V_1$ the parameters in the region $z<0$ and as $g_2$, $V_2$ the parameters in the region $z>0$. For simplicity, they are fine-tuned so that a globally uniform stationary solution exists.

\subsection*{Homogeneous case}
\label{homo}

Many properties of the solutions can be derived in the homogeneous case without discontinuity. We thus momentarily assume $g$ and $\mu$ are uniform and discuss the solutions of \eq{eq:f}. We begin with the homogeneous solutions. Setting $f''=0$ in (\ref{eq:f}) gives
\be \label{poly1}
  2 g f^6-2 \mu  f^4+ J^2=0 .
\ee
\eq{poly1} is a third-order polynomial in $f^2$, which can be solved exactly. There are obviously no real solutions in $f$ for $\mu \leq 0$. We therefore assume $\mu > 0$. A straightforward calculation shows that there are then real solutions if \be \label{eq:Jmax} |J| \leq J_{\max }=\sqrt{\frac{8}{27}\frac{\mu^3}{g^2}} .
\ee
We assume this condition is satisfied. For $|J| < J_{\max }$ there are two homogeneous solutions. It is easily seen that one of them is supersonic while the other one is subsonic. 

It is convenient to treat \eq{eq:f} as a system of coupled first-order equations for $f,p$ with $p \equiv f'$
\be 
\left(
\begin{array}{c}
 f \\
 p
\end{array}
\right)'=\left(
\begin{array}{c}
 p \\
  -2 \mu  f + 2 g f^3+\frac{J^2}{f^3}
\end{array}
\right)
\ee 
Each stationary solution draws a trajectory in phase space $(f,p)$. 
As can be seen in the left panel of Fig.~\ref{phase_portrait2}, they divide the phase space into three regions. The two external ones correspond to solutions which go to infinity at a finite distance from the origin, so they cannot be solutions in an infinite or semi-infinite interval. They also turn out to be irrelevant for the black hole laser case despite the presence of a finite supersonic one. For this reason we will not consider them.
We will instead focus on solutions in the middle domain and at the boundaries.
The former contains periodic solutions which oscillate around the supersonic homogeneous 
one. Their wavelength varies between a finite value in the limit of small amplitudes (solutions which remain close to the homogeneous supersonic one) and infinity close to the boundary. The minimum wavelength is given by \eq{eq:lambda}. The boundary of this region is the dark soliton. At the boundary between the two external domains one finds solutions which are asymptotically divergent on one side and go to a constant on the other side. 

\subsection*{One horizon: The $2+1$ inhomogeneous solutions.}
\label{2+1sol}

When taking the discontinuity into account, we have two phase diagrams: one for $z<0$ and one for $z>0$. There are \textit{a priori} many qualitatively distinct global solutions. But only few of them are relevant for the problem at hands. An important technical simplification comes from the assumption that there exists a globally homogeneous solution. We want this solution to be subsonic for $z<0$ and supersonic for $z>0$. Depending on the sign of the velocity, this is either a 
black- or white-hole horizon. Since this sign does not affect the stationary solutions, we will not specify it. Our analysis will thus be directly applicable to the black hole laser case since the second 
discontinuity basically duplicates the analysis.

Keeping in mind the black hole laser case, we are interested in solutions that go to the subsonic homogeneous solution as $z \rightarrow - \infty$. 
The qualitative properties of the solutions can be seen by superimposing the two phase portraits associated with the two choices of parameters (see \ref{phase_portrait2}, right panel). Relevant solutions start on the black dot at $z \rightarrow - \infty$ (this is equivalent to saying that they go to $f_0$ at $-\infty$). 
There are four possibilities: 

\hspace*{0.5 cm} *the homogeneous  solution with $f=f_0$;

\hspace*{0.5 cm} *the solution following the black line with increasing $f$ and $p$;

\hspace*{0.5 cm} *or the two solutions with (initially) decreasing $f$ and $p$.

In any case the black line 
 is followed until $z=0$. Then the trajectory changes and follows a blue line in Fig.~\ref{phase_portrait2}. If the solution was homogeneous for $z<0$, it remains so for $z>0$. Other solutions are periodic for $z>0$: $f$ shows oscillations around $f_0$ with a finite amplitude $a \equiv (f_{max}-f_0)/f_0$. In the limit of small amplitudes, the wave vector is then given by the nonvanishing root of \eq{eq:disprel1} with $\om = 0$. The wave length goes to infinity when the maximum value of $f$ approaches the subsonic solution $\fb$ given by 
\be \label{eq:fb} 
\fb = \frac{1}{2}\frac{|v|}{c_2} \sqrt{1+\sqrt{1+8\left(\frac{c_2^2}{v^2}\right)}} .
\ee

\begin{figure}
\includegraphics[scale=0.12]{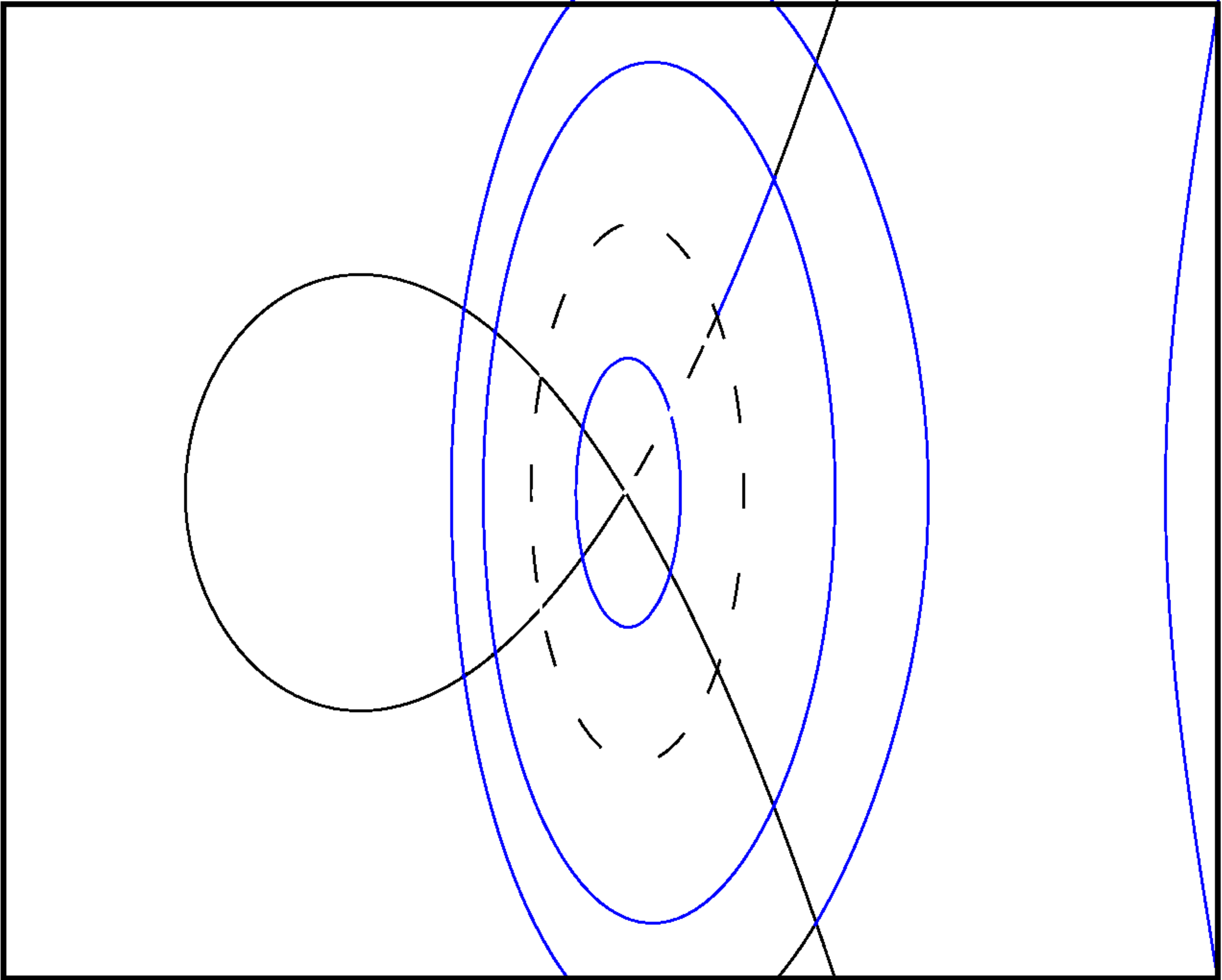}
\includegraphics[scale=0.12]{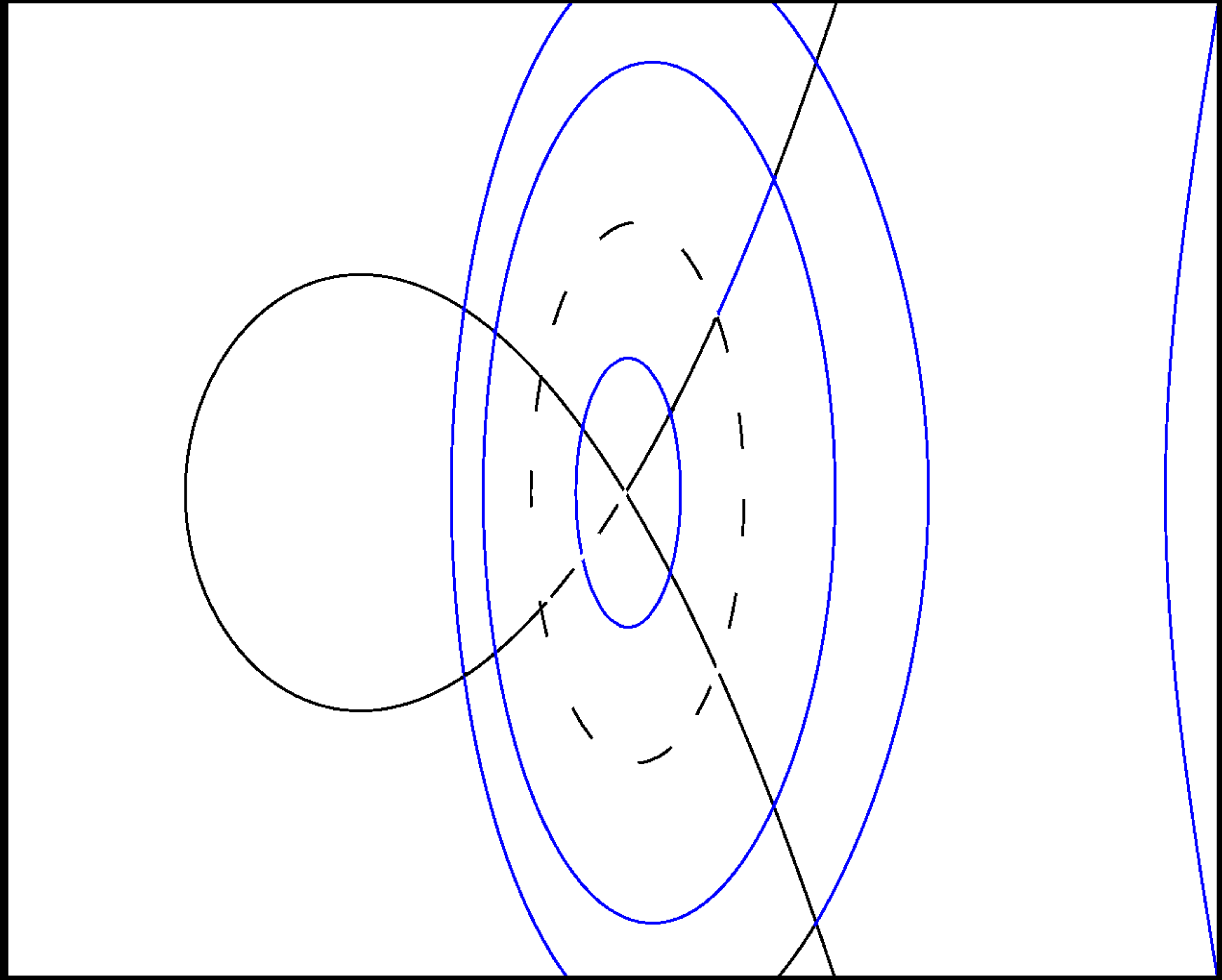}
\includegraphics[scale=0.12]{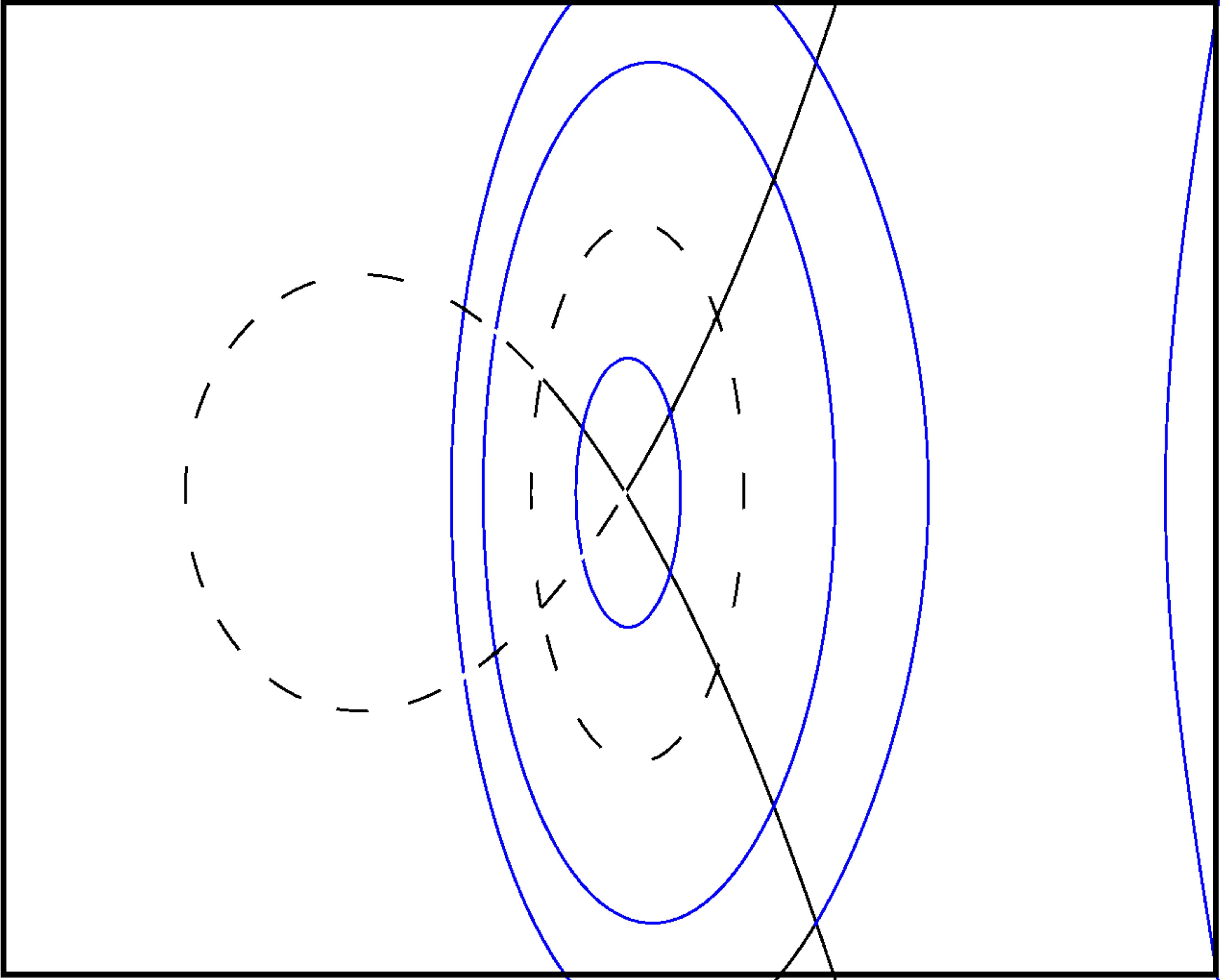}
\includegraphics[scale=0.6]{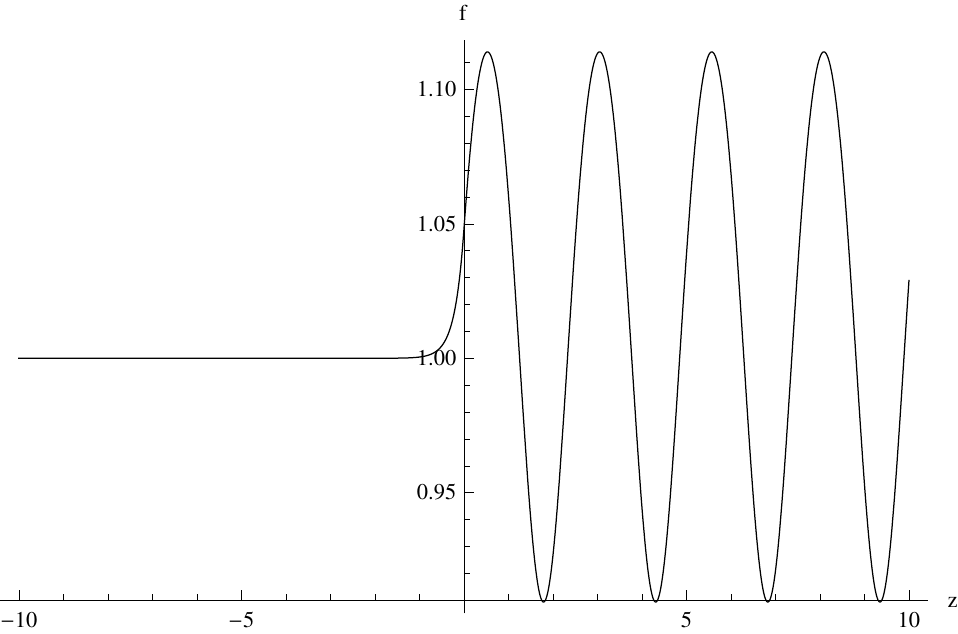}
\includegraphics[scale=0.6]{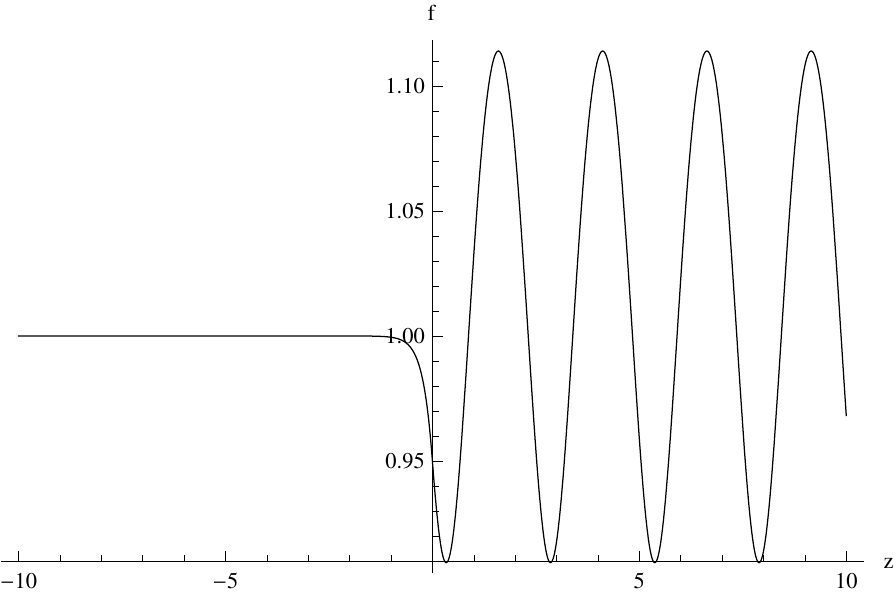}
\includegraphics[scale=0.6]{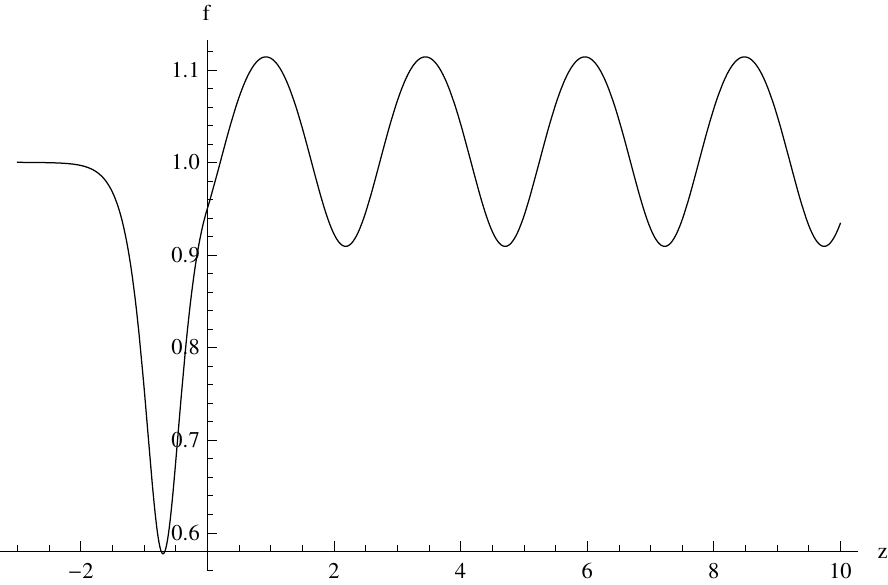}
\caption{Trajectories in phase space (top) and $f$ as a function of $z$ (bottom) for $a \approx 0.115$. Top: Simplified phase portrait where only three blue lines and the black line of Fig.~\ref{phase_portrait2} are represented. The dashed curve corresponds to a solution going to $f=f_0$ at $z=- \infty$. We show the three solutions giving the same amplitude in the region $z>0$. The parameters are $g_1=8$, $g_2=1$, $\mu_1=28/3$, $\mu_2=7/6$ and $J^2=8/3$. Bottom: $f/f_0$ as a function of $z$ for these three solutions. The soliton can be seen on the right plot for $z<0$. The left plots correspond to the shadow-soliton solution, the middle ones to the first soliton solution and the right ones to the second soliton solution.}\label{fig:2+1}
\end{figure}

The first nonlinear solution is obtained by following the black line in the direction of increasing $f$ and $p$ (we will call this solution the \textit{shadow-soliton solution}).
The last two are found by following it in the direction of decreasing $f$ and $p$ (the black loop in Fig.~\ref{phase_portrait2}). There are two of them because, if the black loop crosses a blue line, it does it twice by symmetry $p \rightarrow -p$. The two intersection points give two solutions. If the amplitude of the oscillations is small, one solution corresponds to a very small path on the black loop, hence a tiny fraction of the soliton, while the other one has a nearly complete soliton in the region $z<0$.
The three trajectories in phase space and their corresponding profiles $f(z)$ are represented in Fig.~\ref{fig:2+1}. The second soliton solution is physically less interesting since it has a larger, finite 
energy and cannot be continuously deformed into the homogeneous solution while keeping the oscillations for $z>0$ small. This is the meaning of "$2+1$" in the title of this subsection: For a given (small) amplitude there are three nonlinear solutions, but one of them has a much larger energy than the other two. To linear order, the latter are related by a $\mathbb{Z}_2$ symmetry $\delta f \rightarrow - \delta f $ and correspond to the undulation (zero-frequency wave) described in \cite{CUndulation}.

For larger amplitudes there may be only one solution if the corresponding blue line does not cross the black loop. A straightforward calculation shows the three solutions persist up to the maximum amplitude (at which $f$ reaches $\fb$ asymptotically) if and only if
\be
\frac{2 f_0^4}{\fb^2+f_0^2} \geq  \frac{2 \fp^4}{\fp^2+f_0^2},
\ee 
where $\fp$ is the homogeneous supersonic solution for $z<0$. This can be rewritten as
\be \label{eq:inf}
\frac{v^4}{c_1^2 \, c_2^2}\left(1+\sqrt{1+8 \, \frac{c_2^2}{v^2}}\right)^2 \leq 16,
\ee 
where $c_1$ and $c_2$ are, respectively, the sound velocities for $z<0$ and $z>0$.

The same analysis applies at each horizon of the black hole laser. The decomposition $2+1$ then becomes $(2+1) \times (2+1)=4+5$, \textit{i.e.} $4$ solutions can be 
arbitrarily close to the homogeneous one and can be studied at linear order, while five of them contain at least one soliton. The first four solutions are analogous to those described in~\cite{PhysRevB.53.6693}.

\subsection*{Thermodynamic considerations}
\label{thermo_app}

We use the grand-canonical ensemble: The temperature (set to zero) and chemical potential are fixed while the energy and number of particles depend on the solution. 
We will study two cases: fixed mean velocity or fixed current $J$. 
At fixed velocity\footnote{More precisely, the fixed quantity is the difference in the phases $\theta$ evaluated at two points $z_+ \gg L$ and $z_- \ll -L$.}, 
the (off-shell) energy functional is the grand potential
\be \label{eq:G_off}
G[\psi]=\int \left(\frac{1}{2}\left\lvert \frac{\partial \psi }{\partial  z}\right\rvert^2 -\mu  \left\lvert \psi \right\rvert^2 +\frac{1}{2}g \left\lvert \psi \right\rvert^4 \right) \, dz + C^{te}.
\ee 
It is defined up to a constant, which we choose so that $G=0$ for the global homogeneous solution. Using (\ref{eq:f}), one finds the on-shell function
\be \label{eq:G_on}
G(a,J,i) = \int \frac{dz}{2} \, \lp \frac{d}{dz} \lp f_{a,J,i}(z) f_{a,J,i}'(z) \rp - g(z) \lp f_{a,J,i}(z)^4 - f_0^4 \rp \rp, 
\ee
where $a$ is the amplitude of the solution for $z>0$, $J$ the current, 
and $i$ a discrete parameter 
telling which of the above three solutions $f_{a,J,i}$ is considered. 

The first term in (\ref{eq:G_on}) is a boundary term.
The only contribution comes from $z \rightarrow + \infty$ since 
we assume $f' \rightarrow 0$ at $z \rightarrow - \infty$.
For definiteness we suppose (for a moment) the supersonic region $z>0$ is finite, although arbitrarily large, with a length $l$ equal to an integer number $n$ times the wavelength $\lambda$. In that case the on-shell grand potential reduces to
\be \label{GE}
G = - \int_{-\infty}^{0} \frac{dz}{2} \,   g_1 \lp f(z)^4 - f_0^4 \rp - \int_{0}^{n \, \lambda} \frac{dz}{2} \,   g_2 \lp f(z)^4 - f_0^4 \rp .
\ee 
$G$ can be divided into two contributions. That of the region $z<0$ is finite and comes from the deformation of the solution close to the horizon. The contribution of the region $z>0$ is proportional to $n$ (there is no boundary term because the solution is exactly periodic in this region, without any border effect). The proportionality coefficient, \textit{i.e.} the difference in Gibbs energy per period, is always negative and diverges at the maximum amplitude. 

\eq{eq:G_off} 
is the Gibbs functional in the grand-canonical ensemble if the mean value of the condensate velocity is fixed. This can be seen by computing the total on-shell variation of $G$
\be 
\delta G = -N \delta \mu  +\left[ \delta f \, \partial_z f \right]_{-\infty }^{\infty }+J \left[ \delta \theta \right]_{-\infty }^{\infty },
\ee
where  
$\left[ X \right]_{-\infty }^{\infty } \equiv \text{lim}_{z \rightarrow \infty} \lp X(z) - X(-z) \rp$.
To characterize exact solutions, we found it is more convenient to work at fixed current $J$. The relevant Gibbs energy is then the Legendre transform of (\ref{eq:G_off}) 
\be \label{eq:E}
E \equiv G - \int \, J \pd_z \theta \, dz = \int  \left(\frac{1}{2}f'^2-\frac{J^2}{2f^2}-\mu  f^2+\frac{1}{2}g f^4\right) dz.
\ee 
Since $\int \, J \pd_z \theta \, dz$ is a boundary term, the equations of motion are unchanged.
Up to a constant term chosen so that the energy of the homogeneous solution vanishes, (\ref{eq:E}) can be written as
\be \label{eq:E-os} 
\Delta E = \int  \left(-\frac{1}{2}g \lp f^4-f_0^4 \rp - J^2 \lp \frac{1}{f^2} - \frac{1}{f_0^2} \rp \right) dz.
\ee
The contribution of the deformation of the solution in the region $z<0$ to \eq{eq:E} is always positive. Notice also that the change of $E$ per period in the region $z>0$ with respect to the homogeneous solution is {\it third order} in the oscillation amplitude. The reason is that the first- and second-order terms in the expansion of 
\be \label{eq:Ebis}
\int_{z > 0}  \left(\frac{1}{2}f'^2-\frac{J^2}{2f^2}-\mu  f^2+\frac{1}{2}g f^4\right) dz
\ee 
in $f/f_0 - 1$ vanish for $f/f_0 - 1 \propto \cos(k z)$ with $k = 2 \sqrt{v^2 - c_2^2}$. This property is directly related to the stationarity assumption, as we now show. Let us write the condensate wave function as $\psi = \psi_0 + \phi_\om$, where $\psi_0$ is the homogeneous solution 
and $\phi_\om$ is a perturbation, a solution of the Bogoliubov-de~Gennes equation with frequency $\om$. 
Then the first-order term in \eq{eq:Ebis} automatically vanishes and the second-order term is:
\[
E_2 = \int \, \om \, \phi^*(t,z) \, \phi(t,z) \, dz .
\]
In particular, it vanishes if $\om = 0$.

\subsection*{Characterization of the solutions for $z>0$}
\label{char_sol}

\begin{figure}
\begin{center}
\includegraphics[scale=1.0]{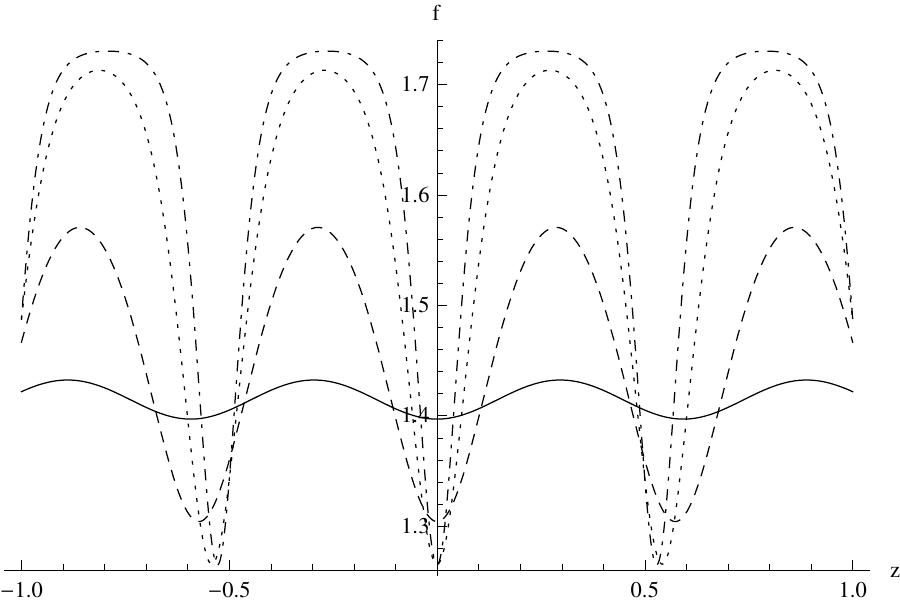}
\caption{Evolution of the shape of the solutions in the region $z>0$ when the wavelength is increased. The parameters are: $g_1=27$, $g_2=5$, $\mu_1 = 63$, $\mu_2 = 19$, $J=6 \sqrt{2}$. The coordinate $z$ is rescaled for each curve so that $z \in [-1,1]$ corresponds to a fixed number of periods. The values of $\lambda / \lambda_0 - 1$ (where $\lambda_0$ is the minimum wavelength for periodic solutions) are $0.0015$ (solid line), $0.11$ (dashed line),  $1.1$ (dotted line)  and $1.9$ (dot-dashed line).}\label{im:NL_sol}
\end{center}
\end{figure}

In the previous subsections we characterized a solution in the region $z>0$ by its amplitude $a=(f_{max}-f_0)/f_0$. This definition is justified because, for large amplitudes, the solution remains mostly close to $f_{max}$; seen Fig.~\ref{im:NL_sol}. In order to get a better understanding of these solutions, here we relate the profile to the wavelength.  

In general solutions of \eq{eq:f} are Weierstrass elliptic functions, with a complex argument for the periodic ones
 \cite{belokolos1994algebro}. Periodic ones can also be expressed in terms of Jacobi elliptic functions \cite{Kamchatnov}.
 Close to the minimum wavelength 
of \eq{eq:lambda}, a straightforward calculation gives the following development for the amplitude:
\be \label{eq:avsla}
\frac{f_{\max }-f_0}{f_0} = 2\left(A+\frac{v^2+c_2^2}{v^2-c_2^2}A^2\right) + \mathcal{O} \lp A^3 \rp,
\ee
where 
\be 
A\equiv \frac{1}{\sqrt{3}}\frac{v^2-c_2^2}{c_2 \sqrt{4v^2+c_2^2}}\sqrt{\frac{\lambda }{\lambda_0}-1}.
\ee
The minimum value of $f$ for a given solution, $f_{min}$, 
is related to $f_{max}$ by 
\be 
f_{\min }^2 &=& 
\frac{2\mu _2f_{\max }^2 -g_2f_{\max}^4-\sqrt{\left(g_2f_{\max}^4 -2\mu _2f_{\max}^2\right)^2-4g_2J^2f_{\max}^2}}{2 g_2f_{\max}^2}  
.
\ee 

We represent the profile of the solutions in Fig.~\ref{im:NL_sol} for increasing values of their amplitudes. 
For small amplitudes $f$ is very close to a sinusoid, in accordance with the results of Sec.~\ref{Slt} and Appendix \ref{App:M-matrix}. To linear order, the $\mathbb{Z}_2$ symmetry $f \rightarrow 2 f_0 - f$ sends the shadow soliton solution to the first soliton solution and \textit{vice-versa}.
This invariance is broken at nonlinear orders as can be seen by the fact that solutions "spend more time" close to $f_{max}$ in order to minimize the energy $E$ by approaching the subsonic density of \eq{eq:fb}.

\section{Evaluation of $E$, $G$ and $L$} 
\label{App:eqs_G_L}

For the interested readers, we give the formulas for the on-shell thermodynamic function 
$G$, and relate the distance between the discontinuities $2 L$ to the integration constants characterizing the solutions. For definiteness we assume $c_1 = c_3$. The generalization to $c_1 \neq c_3$ is straightforward but makes the expressions longer. Because of the discontinuities we must choose three integration constants: $C_1$ in $I_1$, $C_2$ in $I_2$ and $C_3$ in $I_3$. The requirement that the solution goes to $f=f_0$ as $z \rightarrow \pm \infty$ imposes $C_1=C_3 =\lp 2 v^2 + c_1^2 \rp f_0^2$. 
We define
\be \label{finter}
f_{\text{inter},\pm} \equiv f_0\sqrt{1\pm \sqrt{1-\frac{C_1-C_2}{f_0^2\left(c_1^2-c_2^2\right)}}},
\ee
which corresponds to the two possible values of $f$ at $z=\pm L$, and 
\be 
f_s \equiv \frac{v}{c_1} f_0 , 
\ee
which gives the value of $f$ at the bottom of the soliton in $I_1$ or $I_3$. The minimum and maximum values of $f$ for a periodic solution in $I_2$, denoted $f_{min}$ and $f_{max}$, are the first and second positive roots of the polynomial
\be 
g_2 \, X^6 - 2 \mu_2 \, X^4 + C_2 \, X^2 - J^2 = 0 .
\ee
We also define the two functions $p_1$ and $p_2$ as:
\be 
p_i(f) \equiv \frac{1}{f} \sqrt{g_i \, f^6-2 \mu_i \, f^4+C_i f^2-J^2},
\ee
where $i \in \left\lbrace 1,2 \right\rbrace$.

Using these definitions, the expressions of $G$ and $L$ for the nine types of solutions of Fig.~\ref{fig:9-traj} are given below. The corresponding value of $E$ can be deduced from:
\be \label{eq:GtoE}
E = G - \int \frac{df}{p_j(f)} J^2 \lp f^{-2} - f_0^{-2} \rp df .
\ee

\underline{Type 1:}
\be \label{eqs-num-beg}
L=\int _{f_{\text{inter},+}}^{f_{\max }}\frac{df }{p_2(f)}+ n \int _{f_{\min }}^{f_{\max }}\frac{df }{p_2(f)},
\ee
\be 
G=-g_1\int _{f_0}^{f_{\text{inter,+}}}\frac{df }{p_1(f)}\left(f^4-f_0^4\right)-g_2\int _{f_{\text{inter,+}}}^{f_{\max }}\frac{df }{p_2(f)}\left(f^4-f_0^4\right)-n g_2\int _{f_{\min }}^{f_{\max }}\frac{df }{p_2(f)}\left(f^4-f_0^4\right).
\ee

\underline{Type 3:}
\be 
L=-\int _{f_{\text{inter},-}}^{f_{\min }}\frac{df }{p_2(f)}+ n \int _{f_{\min }}^{f_{\max }}\frac{df }{p_2(f)},
\ee
\be 
G=g_1\int _{f_0}^{f_{\text{inter},-}}\frac{df }{p_1(f)}\left(f^4-f_0^4\right)+g_2\int _{f_{\text{inter},-}}^{f_{\min }}\frac{df }{p_2(f)}\left(f^4-f_0^4\right)-n g_2\int _{f_{\min }}^{f_{\max }}\frac{df }{p_2(f)}\left(f^4-f_0^4\right).
\ee

\underline{Types 2 and 4:}
\be 
L=\frac{1}{2}\int _{f_{\text{inter},+}}^{f_{\max }}\frac{df }{p_2(f)}+\frac{1}{2}\int _{f_{\min }}^{f_{\text{inter},-}}\frac{df }{p_2(f)}+2 \lp n+\frac{1}{2}\right) \int _{f_{\min }}^{f_{\max }}\frac{df }{p_2(f)},
\ee
\be 
G&=&-\frac{g_1}{2}\int _{f_{\text{inter},-}}^{f_{\text{inter},+}}\frac{df }{p_1(f)}\left(f^4-f_0^4\right)- \frac{g_2}{2}\int _{f_{\text{inter},+}}^{f_{\max }}\frac{df }{p_2(f)}\left(f^4-f_0{}^4\right)\nn &&-\frac{g_2}{2}\int _{f_{\min }}^{f_{\text{inter},-}}\frac{df }{p_2(f)}\left(f^4-f_0^4\right)-\lp n+\frac{1}{2} \rp g_2\int _{f_{\min }}^{f_{\max }}\frac{df }{p_2(f)}\left(f^4-f_0^4\right).
\ee

\underline{Types 5 and 7:}
\be 
L= (n+1) \int _{f_{\min }}^{f_{\max }}\frac{df }{p_2(f)},
\ee
\be 
G=g_1\int _{f_0}^{f_s}\frac{df }{p_1(f)}\left(f^4-f_0^4\right)- (n+1) g_2\int _{f_{\min }}^{f_{\max }}\frac{df }{p_2(f)}\left(f^4-f_0^4\right).
\ee

\underline{Types 6 and 8:}
\be 
L=\frac{1}{2}\int _{f_{\text{inter},+}}^{f_{\max }}\frac{df }{p_2(f)}+\frac{1}{2}\int _{f_{\text{inter},-}}^{f_{\max }}\frac{df }{p_2(f)}+ n \int _{f_{\min }}^{f_{\max }}\frac{df }{p_2(f)},
\ee
\be 
G&=&-\frac{g_1}{2}\int _{f_s}^{f_{\text{inter},+}}\frac{df }{p_1(f)}\left(f^4-f_0{}^4\right)-\frac{g_1}{2}\int _{f_s}^{f_{\text{inter},-}}\frac{df }{p_1(f)}\left(f^4-f_0^4\right)-\frac{g_2}{2}\int _{f_{\text{inter},+}}^{f_{\max }}\frac{df }{p_2(f)}\left(f^4-f_0^4\right) \nn &&
-\frac{g_2}{2}\int _{f_{\text{inter},-}}^{f_{\max }}\frac{df }{p_2(f)}\left(f^4-f_0^4\right)-ng_2\int _{f_{\min }}^{f_{\max }}\frac{df }{p_2(f)}\left(f^4-f_0^4\right).
\ee

\begin{figure}
\begin{center}
\includegraphics[scale=0.5]{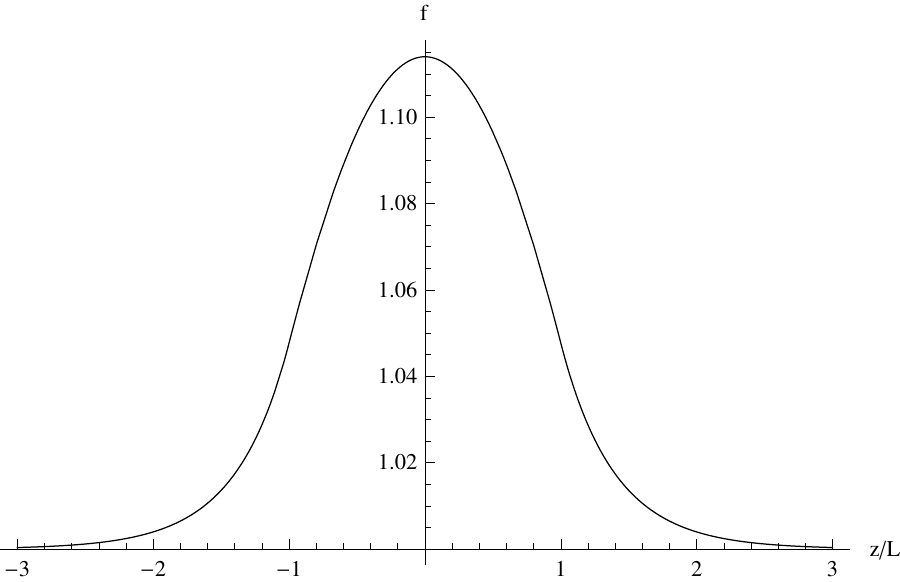}
\includegraphics[scale=0.5]{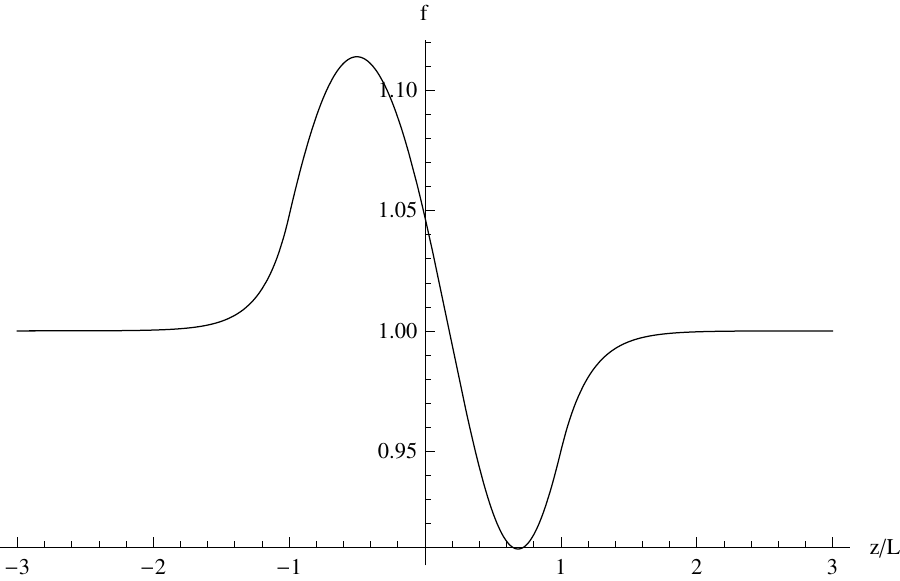}
\includegraphics[scale=0.5]{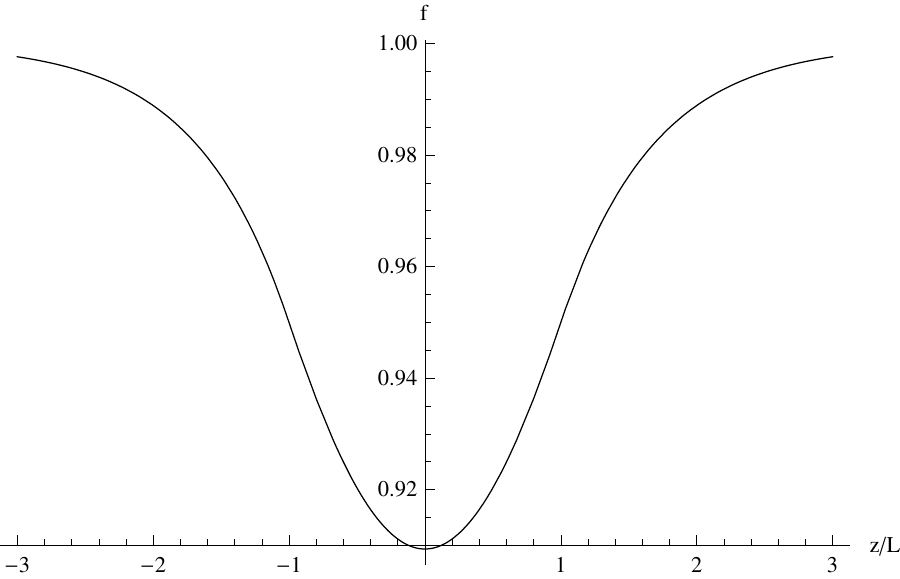}
\includegraphics[scale=0.5]{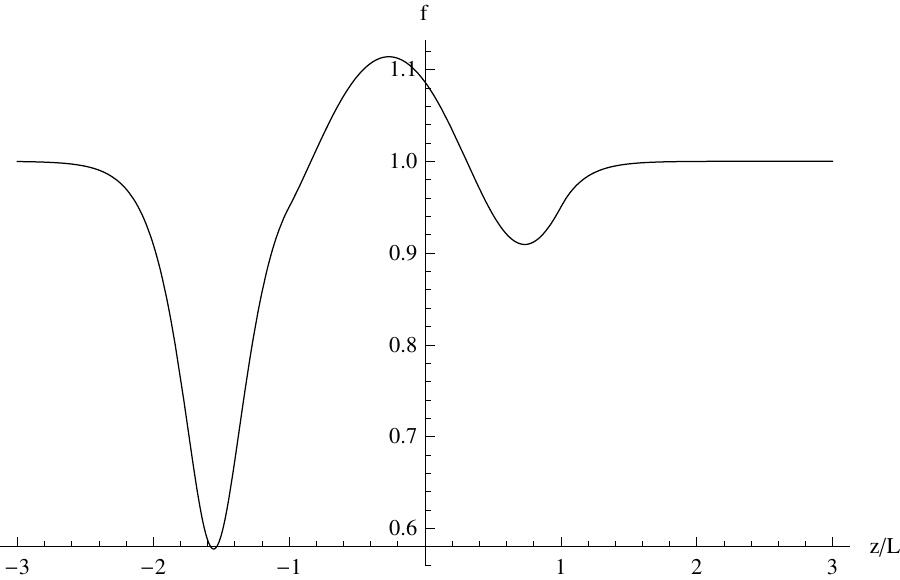}
\includegraphics[scale=0.5]{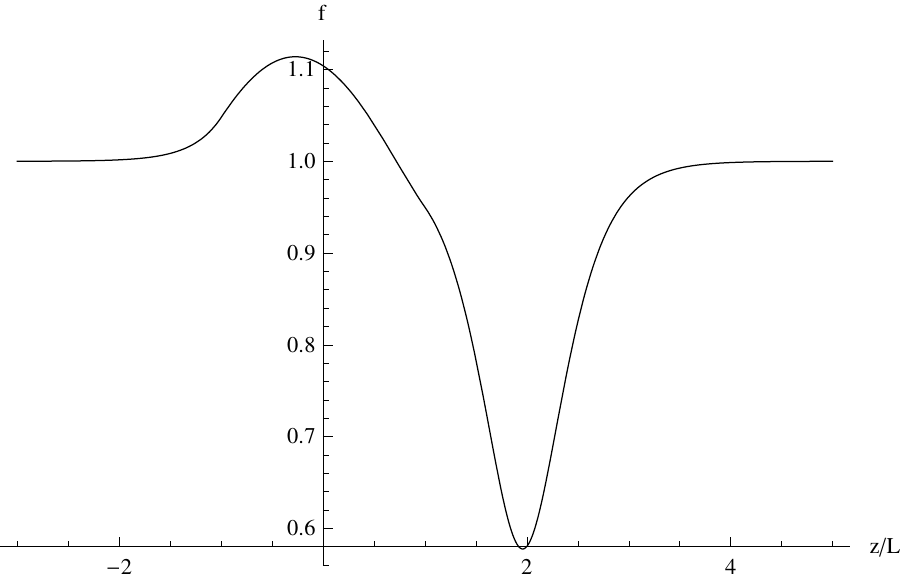}
\includegraphics[scale=0.5]{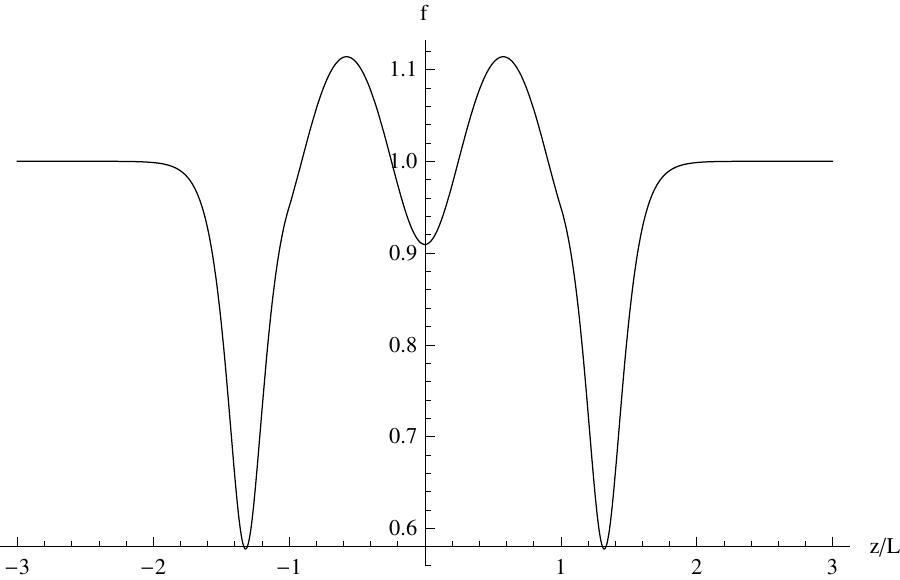}
\caption{$f$ as a function of $z/L$ for solutions of type 1 (top,left), type 2 (top, middle), type 3 (top,right), type 5 (bottom, left), type 6 (bottom, middle) and type 9 (bottom, right) with $n=0$. The parameters are $c_1 = 2 \sqrt{2}$, $c_2 = 1$, $v=\sqrt{8/3}$, $f_0=1$ and $C_2=6.4$. Solutions of types 4, 7 and 8 are obtained from those of types 2, 5 and 6 by the symmetry $z \rightarrow -z$.
}\label{fig:profiles}
\end{center}
\end{figure}

\underline{Type 9}
\be 
L=\int _{f_{\text{inter},-}}^{f_{\max }}\frac{df }{p_2(f)}+ n \int _{f_{\min }}^{f_{\max }}\frac{df }{p_2(f)},
\ee
\be \label{eqs-num-end}
G=g_1\int _{f_0}^{f_s}\frac{df }{p_1(f)}\left(f^4-f_0^4\right)-g_1\int _{f_s}^{f_{\text{inter},-}}\frac{df }{p_1(f)}\left(f^4-f_0^4\right)- g_2\int _{f_{\text{inter},-}}^{f_{\max }}\frac{df }{p_2(f)}\left(f^4-f_0^4\right)-n g_2\int _{f_{\min }}^{f_{\max }}\frac{df }{p_2(f)}\left(f^4-f_0^4\right).
\ee
The integration constant $C_2$ can take any value between $C_{2,min}$ and $C_{2,max}$. $C_{2,min}$ is always equal to:
\be \label{eq:C2min} 
C_{2,min} = \left(2 v^2+c_2{}^2\right) f_0^2 .
\ee  
In general, the maximum value of $C_2$ is:
\be \label{eq:C2max}
C_{2,max}=\frac{v^2}{8 c_2^2}\left(-4c_2^2+v^2+v^2\left(1+8 \frac{c_2^2}{v^2}\right)^{3/2}\right) .
\ee
The only exception is type 3 with $n=0$, for which
\be 
C_{2,\max (3,n=0)}=\left(\frac{c_1^6+2 c_1^2c_2{}^2v^2-v^4c_2{}^2+c_1{}^2v^4}{c_1^4}\right)f_0^2 .
\ee
Figure~\ref{fig:profiles} show $f$ as a function of $z/L$ for the nine different types of solutions and $n=0$. 

\bibliographystyle{h-physrev}
\bibliography{biblio}

\end{document}